\documentclass[sts,preprint]{imsart}    
\usepackage{fullpage}
\usepackage{graphicx}
\usepackage{amsmath}
\usepackage{natbib}
\usepackage{mathrsfs}
\usepackage{amssymb}
\usepackage{subfigure}
\usepackage{comment}

\usepackage{xcolor}

\newcommand{\iid}{\stackrel{\rm iid}{\sim}}
\newcommand{\ind}{\stackrel{\rm ind}{\sim}}

\renewcommand{\th}{\theta}

\newcommand{\eps}{\epsilon}

\newcommand{\cor}{\mathrm{Corr}}

\newcommand{\tht}{\widetilde{\theta}}

\newcommand{\ths}{\th^\star}

\newcommand{\Gs}{G^\star}

\newcommand{\bm}[1]{\boldsymbol{#1}}
\newcommand{\bsc}[1]{\mathbb{#1}}

\renewcommand{\sb}{\bm{s}}
\newcommand{\xb}{\bm{x}}
\newcommand{\yb}{\bm{y}}

\newcommand{\zb}{\bm{z}}
\newcommand{\bfbeta}{\bm{\beta}}
\newcommand{\etab}{\bm{\eta}}
\newcommand{\thb}{\bm{\th}}
\newcommand{\mub}{\bm{\mu}}
\newcommand{\betab}{\bm{\beta}}

\newcommand{\XX}{\mathscr{X}}
\newcommand{\FF}{\mathscr{F}}
\newcommand{\GG}{{\mathcal G}}

\newcommand{\YY}{\mathscr{Y}}
\newcommand{\RR}{\mathscr{R}}
\newcommand{\Psib}{\bm{\Psi}}
\newcommand{\Sigb}{\bm{\Sigma}}
\newcommand{\btheta}{\boldsymbol{\theta}}
\newcommand{\bbeta}{\boldsymbol{\beta}}
\newcommand{\bphi}{\boldsymbol{\phi}}

\newcommand{\DP}{\mbox{DP}}

\newcommand{\normal}{\mbox{N}}

\newcommand{\Be}{\mbox{Be}}

\endlocaldefs

\makeatletter
\def\moverlay{\mathpalette\mov@rlay}
\def\mov@rlay#1#2{\leavevmode\vtop{%
   \baselineskip\z@skip \lineskiplimit-\maxdimen
   \ialign{\hfil$\m@th#1##$\hfil\cr#2\crcr}}}
\newcommand{\charfusion}[3][\mathord]{
    #1{\ifx#1\mathop\vphantom{#2}\fi
        \mathpalette\mov@rlay{#2\cr#3}
      }
    \ifx#1\mathop\expandafter\displaylimits\fi}
\makeatother

\newcommand{\bigcupdot}{\charfusion[\mathop]{\bigcup}{\bullet}}

\begin{document}

\begin{frontmatter}
\title{The Dependent Dirichlet Process and Related Models}
\runtitle{Dependent Dirichlet Processes}

\begin{aug}
\author{\fnms{Fernand A.} \snm{Quintana}\thanksref{m1, m2}\ead[label=e1]{quintana@mat.uc.cl}}
\author{\fnms{Peter} \snm{M\"uller}\thanksref{m3}\ead[label=e2]{pmueller@math.utexas.edu}},
\author{\fnms{Alejandro} \snm{Jara}\thanksref{m1,m2}\ead[label=e3]{atjara@uc.cl}}
\and
\author{\fnms{Steven N.} \snm{MacEachern}\thanksref{m4}\ead[label=e4]{snm@stat.osu.edu}}

\runauthor{Quintana, M\"uller, Jara and MacEachern}

\affiliation{Pontificia Universidad Cat\'olica de Chile \thanksmark{m1} and Millennium Nucleus Center
for the Discovery of Structures in Complex Data \thanksmark{m2} and The University of Texas at Austin \thanksmark{m3}
and The Ohio State University \thanksmark{m4}}

\address{Departamento de Estad\'istica, \\Santiago, Chile\\
\printead{e1} \printead{e3}}

\address{Department of Statistics and Data Science, \\Austin, Texas\\
\printead{e2}}

\address{Department of Statistics, \\Columbus, Ohio \\
\printead{e4}}
\end{aug}

\begin{abstract}
Standard regression approaches assume that some finite number of the
response distribution characteristics, such as location and scale, change
as a (parametric or nonparametric) function of predictors. However, it is
not always appropriate to assume a location/scale representation, where
the error distribution has unchanging shape over the predictor space. In
fact, it often happens in applied research that the distribution of
responses under study changes with predictors in ways that cannot be
reasonably represented by a finite dimensional functional form. This can
seriously affect the answers to the scientific questions of interest, and
therefore more general approaches are indeed needed. This gives rise to
the study of fully nonparametric regression models. We review some of the
main Bayesian approaches that have been employed to define probability
models where the complete response distribution may vary flexibly with
predictors. We focus on developments based on modifications of the
Dirichlet process, historically termed dependent Dirichlet processes, and
some of the extensions that have been proposed to tackle this general
problem using nonparametric approaches. \\

\end{abstract}


\end{frontmatter}

\section{Introduction}\label{sect:intro}

We review the popular class of dependent Dirichlet process (DDP) models.
These define a widely used fully nonparametric Bayesian regression for a
response $\yb \in \YY$, based on a set of predictors $\xb \in \XX
\subseteq \mathbb{R}^p$. Despite a barrage of related literature over the
past 25 years, to date there is no good review of such models. This paper
fills this gap.

Fully nonparametric regression can be seen as an extension of traditional
regression models, where, starting from some elements in $\XX$ and a
corresponding set of responses in $\YY$, the goal is to model the
distribution of $\yb$ given $\xb$. Standard linear regression models
proceed under the assumption of a Gaussian distribution for $\yb\mid\xb$
with a mean modeled as a linear combination of $\xb$. Further extensions
of this idea to exponential families gave rise to the popular class of
generalized linear models, where a transformation of the mean response is
modeled as a linear combination of $\xb$. Many other similar extensions
are available. We focus on a nonparametric version of this idea, which
involves going beyond the notion that the effect of predictors is
restricted to change some particular functional of the response
distribution, such as the mean, a quantile, or the parameters in a
generalized linear model.

The fully nonparametric regression problem that we focus on arises when we
assume that $\yb_i \mid F_{\xb_i} \overset{ind.}{\sim} F_{\xb_i}$,
$i=1,\ldots,n$. The parameter of interest is the complete set of
predictor-dependent random probability measures $\FF=\{F_{\xb} : \xb \in
\XX\}$, where $F_{\xb}$ is a probability measure defined on the response
sample space $\YY$, whose elements can flexibly change with the values of
the predictors $\xb$, i.e. the entire shape of the distribution can change
with $\xb$. From a Bayesian point of view, the fully nonparametric
regression model is completed by defining a prior distribution for $\FF$,
which is taken to be the probability law of a probability measure-valued
stochastic process with index $\xb$. At the risk of abusing notation, we
use from now on the same symbol to refer to the probability measure and
its cumulative distribution function (CDF). The distinction should be
clear from the context.

 Several popular approaches have been developed to formalize Bayesian
inference for such nonparametric regression. These include additive random
tree models like the BART \citep{chipman2010}, approaches based on basis
expansions such as wavelet regression and more. Also, there is of course
extensive literature on non-Bayesian approaches to nonparametric
regression. Many of these approaches are based on a
model of the form
$$
\yb_i=f(\xb_i)+\epsilon_i,\qquad i=1,\ldots,n,
$$
with $E(\epsilon_i)=0$, and
are concerned with finding a function $f:\XX\rightarrow \YY$ such
that $\|y_i-f(\xb_i)\|$ is small, for $f$ in some some class, often
represented as being spanned by some basis functions. Such
methods  include the following:
Under {\em local averaging} $f(\xb)$ is estimated from
those $\yb_i$'s such that $\xb_i$ is ``close'' to $\xb$; {\em local
modeling} estimates $f(\xb)$ by locally fitting some function or
kernel such as a Gaussian function or a polynomial; {\em global modeling
or least squares estimation} finds $f$ that minimizes
$\frac{1}{n}\sum_{i=1}\|\yb_i-f(\xb_i)\|^2$ in the class; and
{\em penalized modeling} is based on finding $f$ that minimizes
$\frac{1}{n}\sum_{i=1}\|\yb_i-f(\xb_i)\|^2+J_n(\yb)$ in the class, where
$J_n(f)$ is a penalization term, such as $J_n(f)=\lambda_n\int_{\XX}
|f^{\prime\prime}(t)|^2\, dt$. 
 See, for example, 
\cite{gyorfietal:02,klemela:14,faraway:16} and references within.
Many of these classical
frequentist approaches could be construed to imply nonparametric Bayesian
models, but they are not usually cast as prior probability models for a
family $\FF$ of random probability measures indexed by covariates. 

 In the Bayesian nonparametric (BNP) literature, 
the problem of defining priors over related random probability
distributions has received increasing attention over the past few
years.
To date, most of the BNP priors to account for the dependence of a set of
probability distributions on predictors are generalizations and extensions
of the celebrated Dirichlet process (DP) \citep{ferguson;73,ferguson;74}
and Dirichlet process mixture (DPM) models \citep{lo;84}.
A DPM model  defines a random probability measure as 
\begin{equation}\label{eq:DPM}
  f(\yb \mid G) =\int_{\Theta}
     \psi(\yb,\thb) G(d \thb), \ \ \yb \in \YY,
\end{equation} where $\psi(\bullet,\thb)$ is a continuous density
function, for every $\thb \in \Theta$, and $G$ is a  discrete 
random probability measure with a DP prior.
If $G$ is DP with parameters $(M,G_0)$,
where $M\in \mathbb{R}_0^+$ and $G_0$ is a probability measure on
$\Theta$, written as $G\mid M, G_0 \sim \mbox{DP}(M G_0)$, then the
trajectories of the process can be a.s. represented by the stick-breaking
representation \citep{sethuraman;94}:
\begin{equation}\label{eq:stbrDP}
G(B)=\sum\limits_{h=1}^\infty w_h \delta_{\thb_h}(B),
\end{equation}
where $B$ is any measurable set, $\delta_{\thb}(\cdot)$ is the Dirac
measure at $\thb$, $w_h=V_h \prod\limits_{\ell<h}(1-V_\ell)$, with
$V_h\mid M \overset{iid}{\sim}\Be(1,M)$, $\thb_h \mid G_0
\overset{iid}{\sim} G_0$, and the $\{w_h\}$ and $\{\thb_h\}$ collections
are independent. Discussion of properties and applications of DPs can be
found, for instance, in \cite{mueller;quintana;jara;hanson;2015}. Many BNP
priors for nonparametric regressions $\FF=\{F_{\xb} : \xb \in \XX\}$ are
based on extensions of model \eqref{eq:DPM}. They incorporate dependence
on predictors via the mixing distribution in \eqref{eq:DPM}, by replacing
$G$ with $G_{\xb}$, and the prior specification problem is related to the
modeling of the collection of predictor-dependent mixing probability
measures $\{G_{\xb} : \xb \in \XX\}$.

Consider first the simplest case, where a finite number of dependent RPMs
$\GG=\{G_j,\;j=1,\ldots,J\}$ are judged to be exchangeable so that the
prior model $p(\GG)$ should accordingly be invariant with respect to all
permutations of the indices.
Consider, for example, an application to borrowing strength across $J$
related clincal studies.
This can be achieved, for example, through joint modeling of
study-specific effects distributions $G_j$ for $j=1,\ldots,J$. A main aim
here is that subjects under study $j_1$ should inform inference about
subjects enrolled in a different but related study $j_2 \not= j_1$. Two
extreme modeling choices would be (i) to pool all patients and assume one
common effects distribution, or (ii) to assume $J$ distinct distributions
with independent priors. Formally, the earlier choice assumes $G_j \equiv
G$, $j=1,\ldots,J$, with a prior $p(G)$, such as $G\sim DP(M,G_0)$. The
latter assumes $G_j \sim p(G_j)$, independently, $j=1,\ldots,J$. We refer
to the two choices as extremes since the first choice implies maximum
borrowing of strength, and the other choice implies no borrowing of
strength.  In most applications, the desired level of borrowing strength
is somewhere in-between these two extremes.
\begin{figure}[hbtp]
\begin{center}
\begin{tabular}{cc}
  \setlength{\unitlength}{987sp}%
\begingroup\makeatletter\ifx\SetFigFont\undefined%
\gdef\SetFigFont#1#2#3#4#5{%
  \reset@font\fontsize{#1}{#2pt}%
  \fontfamily{#3}\fontseries{#4}\fontshape{#5}%
  \selectfont}%
\fi\endgroup%
\begin{picture}(11695,4826)(244,-4574)
\put(6451,-2236){\makebox(0,0)[lb]{\smash{\SetFigFont{6}{7.2}{\rmdefault}{\mddefault}{\updefault}{\color[rgb]{0,0,0}G}%
}}}
{\color[rgb]{0,0,0}\thinlines
\put(5176,-3886){\circle{1210}}
}%
{\color[rgb]{0,0,0}\put(7126,-3961){\circle{1210}}
}%
{\color[rgb]{0,0,0}\put(9376,-3886){\circle{1210}}
}%
{\color[rgb]{0,0,0}\put(6601,-2086){\circle{1210}}
}%
{\color[rgb]{0,0,0}\put(6601,-361){\circle{1210}}
}%
{\color[rgb]{0,0,0}\put(857,-3948){\circle{1210}}
}%
{\color[rgb]{0,0,0}\put(11326,-3886){\circle{1210}}
}%
{\color[rgb]{0,0,0}\put(6601,-1036){\vector( 0,-1){450}}
}%
{\color[rgb]{0,0,0}\put(6001,-2161){\vector(-4,-1){4764.706}}
}%
{\color[rgb]{0,0,0}\put(6001,-2311){\vector(-3,-1){3420}}
}%
{\color[rgb]{0,0,0}\put(6226,-2686){\vector(-1,-1){712.500}}
}%
{\color[rgb]{0,0,0}\put(6751,-2761){\vector( 1,-3){225}}
}%
{\color[rgb]{0,0,0}\put(7201,-2386){\vector( 3,-1){2362.500}}
}%
{\color[rgb]{0,0,0}\put(7201,-2236){\vector( 3,-1){3667.500}}
}%
\put(6526,-511){\makebox(0,0)[lb]{\smash{\SetFigFont{6}{7.2}{\rmdefault}{\mddefault}{\updefault}{\color[rgb]{0,0,0}$\eta$}%
}}}
\put(1651,-4111){\makebox(0,0)[lb]{\smash{\SetFigFont{6}{7.2}{\rmdefault}{\mddefault}{\updefault}{\color[rgb]{0,0,0}...}%
}}}
\put(6001,-4111){\makebox(0,0)[lb]{\smash{\SetFigFont{6}{7.2}{\rmdefault}{\mddefault}{\updefault}{\color[rgb]{0,0,0}...}%
}}}
\put(10126,-4111){\makebox(0,0)[lb]{\smash{\SetFigFont{6}{7.2}{\rmdefault}{\mddefault}{\updefault}{\color[rgb]{0,0,0}...}%
}}}
\put(601,-4036){\makebox(0,0)[lb]{\smash{\SetFigFont{6}{7.2}{\rmdefault}{\mddefault}{\updefault}{\color[rgb]{0,0,0}y11}%
}}}
\put(2476,-4036){\makebox(0,0)[lb]{\smash{\SetFigFont{6}{7.2}{\rmdefault}{\mddefault}{\updefault}{\color[rgb]{0,0,0}y1n}%
}}}
\put(4876,-4036){\makebox(0,0)[lb]{\smash{\SetFigFont{6}{7.2}{\rmdefault}{\mddefault}{\updefault}{\color[rgb]{0,0,0}y21}%
}}}
\put(6826,-4036){\makebox(0,0)[lb]{\smash{\SetFigFont{6}{7.2}{\rmdefault}{\mddefault}{\updefault}{\color[rgb]{0,0,0}y2n}%
}}}
\put(9076,-4036){\makebox(0,0)[lb]{\smash{\SetFigFont{6}{7.2}{\rmdefault}{\mddefault}{\updefault}{\color[rgb]{0,0,0}y31}%
}}}
\put(11026,-3961){\makebox(0,0)[lb]{\smash{\SetFigFont{6}{7.2}{\rmdefault}{\mddefault}{\updefault}{\color[rgb]{0,0,0}y3n}%
}}}
{\color[rgb]{0,0,0}\put(2776,-3960){\circle{1210}}
}%
\end{picture} &
  \setlength{\unitlength}{987sp}%
\begingroup\makeatletter\ifx\SetFigFont\undefined%
\gdef\SetFigFont#1#2#3#4#5{%
  \reset@font\fontsize{#1}{#2pt}%
  \fontfamily{#3}\fontseries{#4}\fontshape{#5}%
  \selectfont}%
\fi\endgroup%
\begin{picture}(11695,4826)(244,-4574)
\put(11026,-3961){\makebox(0,0)[lb]{\smash{\SetFigFont{6}{7.2}{\rmdefault}{\mddefault}{\updefault}{\color[rgb]{0,0,0}y3n}%
}}}
{\color[rgb]{0,0,0}\thinlines
\put(5176,-3886){\circle{1210}}
}%
{\color[rgb]{0,0,0}\put(7126,-3961){\circle{1210}}
}%
{\color[rgb]{0,0,0}\put(9376,-3886){\circle{1210}}
}%
{\color[rgb]{0,0,0}\put(6601,-2086){\circle{1210}}
}%
{\color[rgb]{0,0,0}\put(6601,-361){\circle{1210}}
}%
{\color[rgb]{0,0,0}\put(857,-3948){\circle{1210}}
}%
{\color[rgb]{0,0,0}\put(2251,-2161){\circle{1210}}
}%
{\color[rgb]{0,0,0}\put(9976,-2161){\circle{1210}}
}%
{\color[rgb]{0,0,0}\put(11326,-3886){\circle{1210}}
}%
{\color[rgb]{0,0,0}\put(6601,-1036){\vector( 0,-1){450}}
}%
{\color[rgb]{0,0,0}\put(6076,-736){\vector(-3,-1){3307.500}}
}%
{\color[rgb]{0,0,0}\put(7051,-811){\vector( 2,-1){2370}}
}%
{\color[rgb]{0,0,0}\put(2176,-2761){\vector(-4,-3){936}}
}%
{\color[rgb]{0,0,0}\put(2326,-2761){\vector( 1,-3){202.500}}
}%
{\color[rgb]{0,0,0}\put(6226,-2686){\vector(-1,-1){712.500}}
}%
{\color[rgb]{0,0,0}\put(6751,-2761){\vector( 1,-3){225}}
}%
{\color[rgb]{0,0,0}\put(9751,-2761){\vector(-1,-2){255}}
}%
{\color[rgb]{0,0,0}\put(10201,-2761){\vector( 1,-1){675}}
}%
\put(6526,-511){\makebox(0,0)[lb]{\smash{\SetFigFont{6}{7.2}{\rmdefault}{\mddefault}{\updefault}{\color[rgb]{0,0,0}$\eta$}%
}}}
\put(1651,-4111){\makebox(0,0)[lb]{\smash{\SetFigFont{6}{7.2}{\rmdefault}{\mddefault}{\updefault}{\color[rgb]{0,0,0}...}%
}}}
\put(6001,-4111){\makebox(0,0)[lb]{\smash{\SetFigFont{6}{7.2}{\rmdefault}{\mddefault}{\updefault}{\color[rgb]{0,0,0}...}%
}}}
\put(10126,-4111){\makebox(0,0)[lb]{\smash{\SetFigFont{6}{7.2}{\rmdefault}{\mddefault}{\updefault}{\color[rgb]{0,0,0}...}%
}}}
\put(1951,-2311){\makebox(0,0)[lb]{\smash{\SetFigFont{6}{7.2}{\rmdefault}{\mddefault}{\updefault}{\color[rgb]{0,0,0}G1}%
}}}
\put(6451,-2236){\makebox(0,0)[lb]{\smash{\SetFigFont{6}{7.2}{\rmdefault}{\mddefault}{\updefault}{\color[rgb]{0,0,0}G2}%
}}}
\put(9676,-2311){\makebox(0,0)[lb]{\smash{\SetFigFont{6}{7.2}{\rmdefault}{\mddefault}{\updefault}{\color[rgb]{0,0,0}G3}%
}}}
\put(601,-4036){\makebox(0,0)[lb]{\smash{\SetFigFont{6}{7.2}{\rmdefault}{\mddefault}{\updefault}{\color[rgb]{0,0,0}y11}%
}}}
\put(2476,-4036){\makebox(0,0)[lb]{\smash{\SetFigFont{6}{7.2}{\rmdefault}{\mddefault}{\updefault}{\color[rgb]{0,0,0}y1n}%
}}}
\put(4876,-4036){\makebox(0,0)[lb]{\smash{\SetFigFont{6}{7.2}{\rmdefault}{\mddefault}{\updefault}{\color[rgb]{0,0,0}y21}%
}}}
\put(6826,-4036){\makebox(0,0)[lb]{\smash{\SetFigFont{6}{7.2}{\rmdefault}{\mddefault}{\updefault}{\color[rgb]{0,0,0}y2n}%
}}}
\put(9076,-4036){\makebox(0,0)[lb]{\smash{\SetFigFont{6}{7.2}{\rmdefault}{\mddefault}{\updefault}{\color[rgb]{0,0,0}y31}%
}}}
{\color[rgb]{0,0,0}\put(2776,-3960){\circle{1210}}
}%
\end{picture} \\
  (a) & (b)
\end{tabular}
\end{center}
\caption{One common RPM $G$ (panel a) versus distinct RPMs $G_j$, independent
across studies (panel b).  Here $\eta$ is a fixed hyperparameter. }
\label{fig:GorGj}
\end{figure}
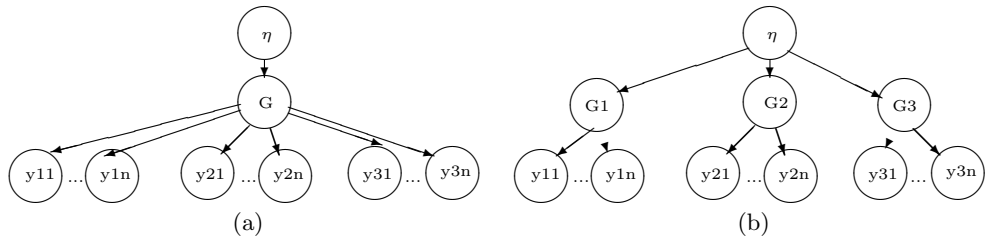

\begin{figure}[hbtp]
\begin{center}
\begin{tabular}{ccc}
\includegraphics[width=.3\textwidth]{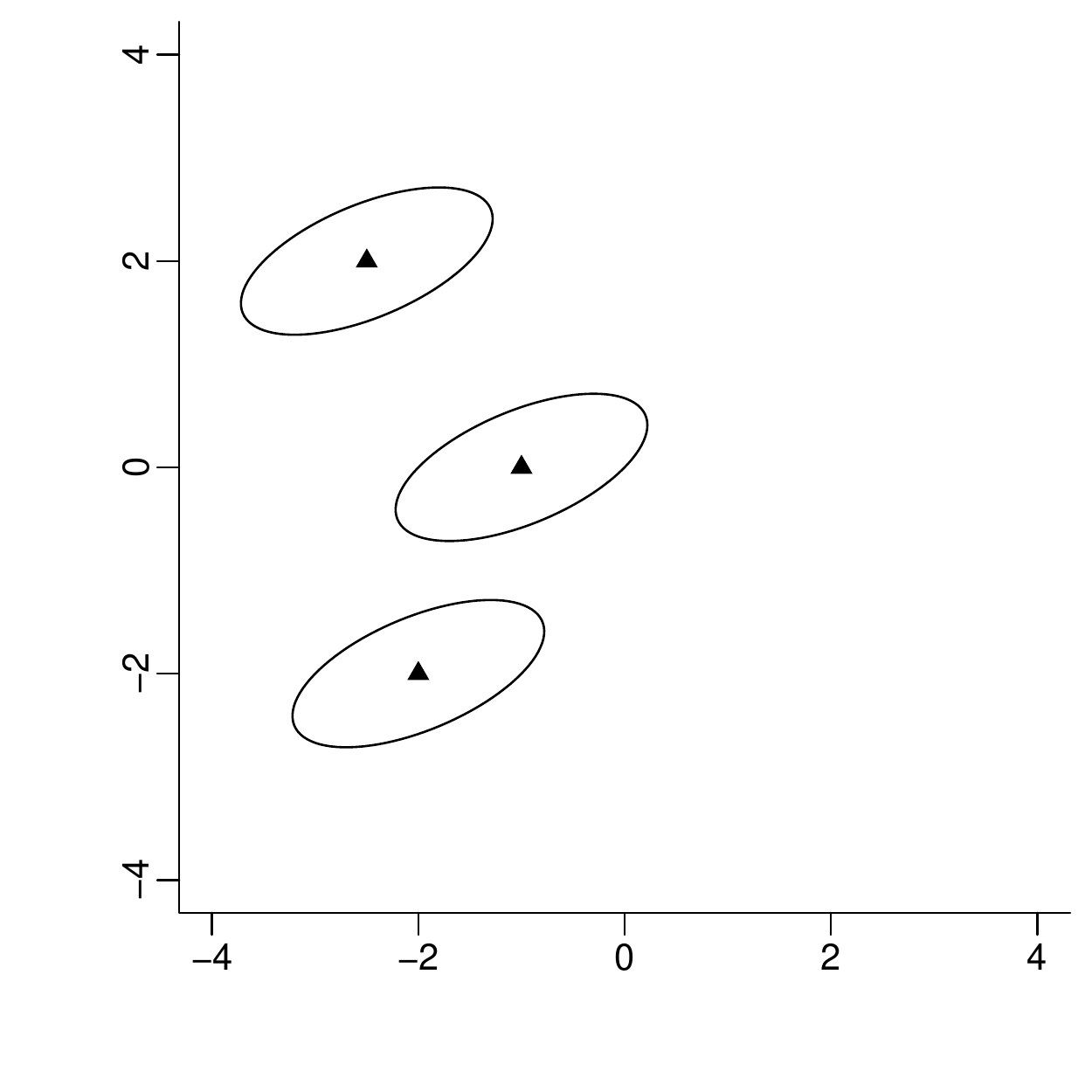} &
\includegraphics[width=.3\textwidth]{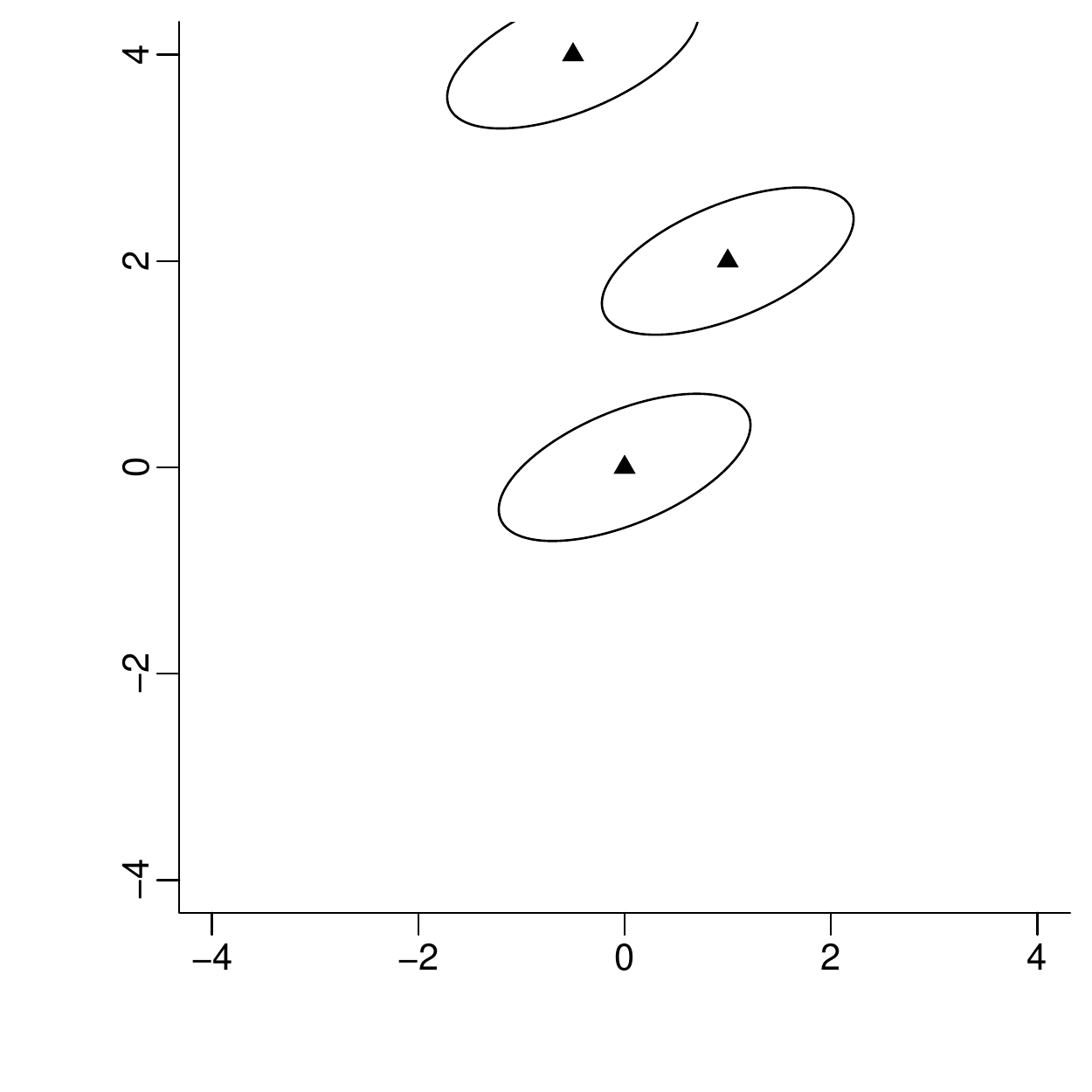} &
\includegraphics[width=.3\textwidth]{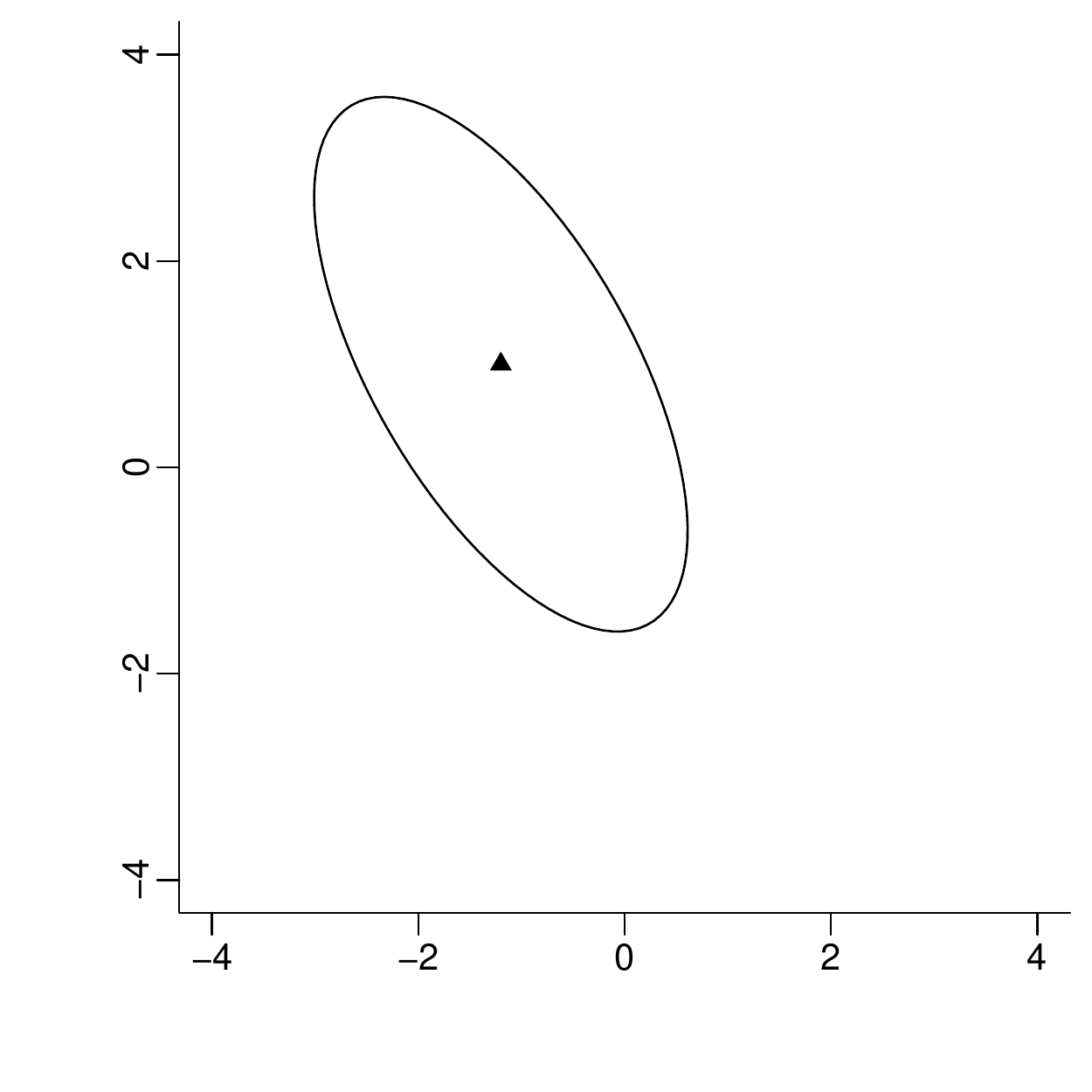}\\
(a) $E(G_1 \mid \yb)$ & (b) $E(G_2 \mid \yb)$ & (c) Prediction for $G_3$
\end{tabular}
\end{center}
\caption{$G_j \sim \DP(M, \Gs)$ with common $\Gs = \normal(m,B)$. Learning across
studies is restricted to the parametric form of $\eta$. The obvious common
structure of $G_1$ and $G_2$ as defining three well separated clusters can
not be learned by the model, which is restricted to learning through the
common hyperparameters $\eta$.} \label{fig:potatoes}
\end{figure}

Figure \ref{fig:GorGj} illustrates the two modeling approaches. Note that
in Figure \ref{fig:GorGj} we added a hyperparameter $\eta$ to index the
prior model $p(G_j \mid \eta)$ and $p(G \mid \eta)$, which was implicitly
assumed fixed. The use of a random hyperparameter $\eta$ allows for some
borrowing of strength even in the case of conditionally independent $p(G_j
\mid \eta)$. Learning across studies can happen through learning about the
hyperparameter $\eta$.
However, the nature of the learning across studies is determined by the
parametric form of $\eta$.  This is illustrated in Figure
\ref{fig:potatoes}.  Assume $G_j \sim \DP(M,\Gs_\eta)$, independently,
$j=1,2,$ and a base measure $\Gs_\eta = \normal(m,B)$ with unknown
hyperparameter $\eta=(m,B)$. In this case, prediction for a future study
$G_3$ can not possibly learn about the multimodality of $G_1$ and $G_2$,
beyond general location and orientation.

%
The previous simple example illustrates the need to develop classes of
models with the ability to relate collections of nonparametric
distributions in more complex fashions. When this collection is indexed by
a set of predictors $\xb\in\XX$, the nonparametric regression approach
mentioned earlier arises, and the definition of a prior on this collection
enables one to borrow information across the distributions for responses,
$F_{\xb}$. For modeling, one important property is the notion of
distributions changing smoothly with respect to $\xb\in\XX$, just as is
the case of generalized linear models in the scale of the transformed
mean. The smoothness could be expressed as continuity of $F_{\xb}$ (with
respect to some conveniently chosen topology) or as the notion that
$F_{\xb}$ ``approaches'' $F_{\xb_0}$ as $\xb\rightarrow \xb_0$, for
instance, $\cor\{{F_{\xb}(A),F_{\xb_0}}(A)\}\rightarrow 1$ as
$\xb\rightarrow \xb_0$ for any event $A$. Many of the models to be
discussed later satisfy some version of this property.

An early reference on predictor-dependent DP models is
\cite{cifarelli;regazzini;78}, who defined a model for related probability
measures by introducing a regression model in the centering measure of a
collection of independent DP random measures.  This approach is used, for
example, by \cite{muliere;petrone;93}, who considered a linear regression
model for the centering distribution of the form $G_{\xb}^0 \equiv
N(\xb'\beta,\sigma^2)$, where $\beta \in \mathbb{R}^p$ is a vector of
regression coefficients, and $N(\mu,\sigma^2)$ stands for a normal
distribution with mean $\mu$ and variance $\sigma^2$.
This is the type of construction illustrated in Figure~\ref{fig:potatoes}.
Similar models were discussed by \cite{mira;petrone;96} and
\cite{giudici;mezzetti;muliere;2003}. Linking the related nonparametric
models through a regression on the baseline parameters of nonparametric
models, however, limits the nature of the trajectories and the type of
dependent processes that can be thus generated. Indeed, realizations of
the resulting process $\GG=\{G_{\xb} : \xb \in \XX\}$ are not continuous
as a function of the predictors. The very limited type of association
structure motivated the development of alternative
 extensions of the DP model to a prior for $\GG$.  In this paper,
we provide an overview of the main constructions of such
predictor-dependent  extensions of DP priors and  and their main
properties. The discussion centers on different ways of constructing the
nonparametric component of models.  A few of the many successful types of
applications that have been proposed are mentioned. In reviewing the
various models to be presented, we discuss some of the main corresponding
works without attempting to provide a complete catalog of references. 
We include a brief discussion of other popular constructions of dependent
DP random measures, without the explicit notion of a conditioning
covariate $x$.

 While we focus on DP-based constructions, we note that several
interesting alternatives to develop predictor-driven random probability
measures have been considered in the recent literature.
\cite{tokdar;zhu;ghosh;2010} develop a logistic Gaussian process that
allows for smoothly varying dependence on conditioning variables. Still
using Gaussian process priors, but starting from a rather different
construction, \cite{jara;hanson;2011} proposed another alternative,
putting the Gaussian process prior on the (logit transformation) of 
the branch   probabilities in a Polya tree prior \citep{lavine;92}.
Another covariate-dependent extension of the Polya tree model was
introduced in \cite{Trippa11} who define a dependent multivariate process
for the  branch   probabilities based on a simple gamma process
construction.  

 Finally,  although issues pertaining to the implementation of
posterior simulation are relevant for practical application of these
methods, our discussion does not focus on computational aspects.

In Section~\ref{sect:DDP} we describe MacEachern's
dependent Dirichlet process (DDP) and its basic properties. In
Section~\ref{variationDDP} we discuss the main variations and alternative
constructions to MacEachern's DDP. In Section~\ref{joint} we discuss
approaches to handle endogenous predictors.  In Section~\ref{sec:part}
we discuss the implied partition structure of DDP models. In
Section~\ref{applications} we illustrate the main approaches. A final
discussion in Section~\ref{sect:disc} concludes the article, including
 some thoughts on future research directions. 

\section{Dependent Dirichlet Process (DDP)}\label{sect:DDP}

We start our discussion with the general definition of DDP and then give
details for popular special cases.

\subsection{General definition}\label{sect:gendef}

\cite{maceachern;99,maceachern;2000} introduced  the DDP model as 
a flexible class of predictor-dependent random probability distributions.
The key idea behind the DDP  construction is to define of a set  
of random measures that are marginally (i.e. for every possible predictor
value $\xb\in\XX$) DP-distributed random measures. In this framework,
dependence is introduced through a modification of the stick--breaking
representation of each element in the set,
\begin{eqnarray}\label{ddp_def}
G_{\xb}(\bullet) = \sum_{h=1}^{\infty}  \underbrace{\left\{V_h(\xb)
\prod_{\ell < h} \left[1 - V_{\ell}(\xb) \right] \right\}}_{w_h(\xb)} \delta_{\btheta_h(\xb)}(\bullet),
\end{eqnarray}
where $V_h(\xb)$, $h \in \mathbb{N}$, are $[0,1]$-valued independent
stochastic processes with index set $\XX$ and $\Be(1,M_{\xb})$ marginal
distributions, and $\btheta_h(\xb)$, $h \in \mathbb{N}$, are independent
stochastic processes with index set $\XX$ and $G^0_{\xb}$ marginal
distributions. The processes associated to the weights and atoms are
independent. From an intuitive viewpoint, the constructed DDP can be
thought of as taking an ordinary DP and modifying some of its components
(i.e. weights and atoms) according to the type of desired indexing or
functional dependence of predictors $\xb\in\XX$. Conditions on the
$V_h(\xb)$ and $\btheta_h(\xb)$ processes can be established to ensure
smoothness of the resulting random measures $G_{\xb}(\bullet)$ when $\xb$
ranges over $\XX $.

\paragraph*{Canonical DDP construction.}
\cite{maceachern;99,maceachern;2000} defined and provided a canonical
construction of the DDP by using transformations of two independent sets
of stochastic processes, $Z^{V_h}_{\XX}=\left \{Z^{V}_h(\xb) : \xb \in
\XX\right\}$, and $Z^{\btheta_h}_{\XX}=\left \{Z^{\btheta}_h(\xb) : \xb
\in \XX\right\}$, for $h\ge 1$, the former used for defining
$\{V_h(\xb)\}$, and the latter for defining $\{\btheta_h(\xb)\}$.
To induce the desired marginal distributions for $\{V_h(\xb)\}$ and
$\{\btheta_h(\xb)\}$, MacEachern resorted to the well-known inverse
transformation method \citep[see, e.g.][]{devroye:1986}.
%
For instance, let $Z(x)$ denote a zero-mean Gaussian process on
$\XX=\bsc{R}$ having constant variance $\sigma^2$. Let $\Phi(\cdot)$ and
$B(\cdot)$ denote the cumulative distribution functions of the $N(0,1)$
and $\Be(1,M)$ distributions, respectively. Then
$V(x)=B^{-1}(\Phi(\sigma^{-1}Z(x)))$ is a stochastic process on $\XX$ that
satisfies $V(x)\sim \Be(1,M)$ for all $x\in\XX$. The same type of
transformation can be applied to construct suitable atom processes
$\{\theta_h(x),\,h\ge 1\}$ such that $\theta_h(x)\sim G_0$ for all
$x\in\XX$ and $h\ge 1$.

Practical application of this general model requires specification of its
various components, which has traditionally motivated the adoption of some
specific forms. The most commonly used DDPs assume that covariate
dependence is introduced either in the atoms or weights, leaving the other
as a collection of random variables exhibiting no covariate indexing, so
that the basic DP definition is partially modified but the distributional
properties retained. We review these forms  in the next section. 

\paragraph*{ Support and an alternative definition.}
One particularity of MacEachern's DDP definition is that given the sets of
stochastic processes, $Z^{V_h}_{\XX}=\left \{Z^{V}_h(\xb) : \xb \in
\XX\right\}$ and $Z^{\btheta_h}_{\XX}=\left \{Z^{\btheta}_h(\xb) : \xb \in
\XX\right\}$, and all other parameters involved in the transformations
described above,
the collection of dependent probability distributions given in
(\ref{ddp_def}) are not random: they are just deterministic functions of
these quantities.
%
To facilitate the  study of theoretical properties of the
DDP,~\cite{barrientos;jara;quintana;2012} gave an alternative definition.
This alternative definition exploits the connection between copulas and
stochastic processes. Since under certain regularity conditions a
stochastic process is completely characterized by its finite-dimensional
distributions, it is possible --and useful-- to define stochastic
processes with given marginal distributions via copulas. The basic idea is
to specify the collection of finite dimensional distributions of a process
through a collection of copulas and marginal distributions.

Copulas are functions that are useful for describing and understanding the
dependence structure between random variables. If $H$ is a $d$--variate
CDF with marginal CDF's given by $F_{1},\ldots,F_{d}$, then by Sklar's
theorem \citep{sklar;59}, there exists a copula function $C:
\left[0,1\right]^{d} \longrightarrow [0,1]$ such that
$H(t_{1},\ldots,t_{d})=C(F_{1}(t_{1}),\ldots,F_{d}(t_{d}))$, for all
$t_{1},\ldots,t_{d}\in\mathbb{R}$, and this representation is unique if
the marginal distributions are absolutely continuous. Thus by the
probability integral transform, a copula function is a $d$--variate CDF on
$\left[0,1\right]^{d}$ with uniform marginals on $\left[0,1\right]$, which
fully captures the dependence among the associated random variables,
irrespective of the marginal distributions.

Let $\mathcal{C}_{\mathcal{X}}^{V}= \left\{
C^V_{\xb_{1},\ldots,\xb_{d}}:\; \xb_{1},\ldots,\xb_{d} \in\mathcal{X},
d>1\right\}$ and $\mathcal{C}_{\XX}^{\theta}=\left\{
C^{\theta}_{\xb_{1},\ldots,\xb_{d}}:\; \xb_{1},\ldots,\xb_{d}
\in\mathcal{X}, d>1\right\}$ be two sets of copulas satisfying
Kolmogorov's consistency conditions.  In
\cite{barrientos;jara;quintana;2012}'s definition, $V_h(\xb)$, $h \in
\mathbb{N}$, are $[0,1]$-valued independent stochastic processes with
index set $\XX$, with common finite dimensional distributions determined
by the set of copulas $\mathcal{C}_{\XX}^{V}$, and $\Be(1,M_{\xb})$
marginal distributions. Similarly, $\btheta_h(\xb)$, $h \in \mathbb{N}$,
are independent stochastic processes with index set $\XX$, with common
finite dimensional distributions determined by the set of copulas
$\mathcal{C}_{\XX}^{\theta}$, and $G^0_{\xb}$ marginal distributions. This
alternative construction produces a definition of the DDP exactly as in
\eqref{ddp_def}, and in particular, the interpretation of the DDP obtained
as modifying a basic DP persists.
%
%
Furthermore, based on this alternative definition,
\cite{barrientos;jara;quintana;2012} established basic properties of
MacEachern's DDP and other dependent-stick breaking processes.
Specifically, they provided sufficient conditions for the full weak
support of different versions of the process and also to ensure smoothness
of trajectories of $G_{\xb}(\bullet)$ as $\xb$ ranges over $\XX$. In
addition, they also characterized the Hellinger and Kullback-Leibler
support of mixtures induced by different versions of the DDP and extended
the results to the general class of dependent stick-breaking processes.

\subsection{The single-weights DDP}\label{sect:swDDP}
MacEachern considered the case of common weights across the values of $\bm{x}$, also referred
to as ``single-weights'' DDP model, defined as
\begin{eqnarray}\label{eq:singleweight}
G_{\xb}(\bullet) = \sum_{h=1}^{\infty}  \underbrace{\left\{V_h
\prod_{\ell < h} \left[1 - V_\ell \right] \right\}}_{w_h} \delta_{\btheta_h(\xb)}(\bullet)=\sum_{h=1}^{\infty}
w_h \delta_{\btheta_h(\xb)}(\bullet),
\end{eqnarray}
where the $V_h$'s are iid $\Be(1,M)$ random variables, which are common
across all levels of $\xb$. The $\btheta_h(\xb)$'s are independent
stochastic processes with index set $\mathcal{X}$ and 
 marginal distributions $G^0_{\xb}$. 
In the literature, to this day, this is the most popular
form of DDP, mainly due to the fact that posterior simulation can be
implemented using the same type of sampling algorithms available for the
case of the DP.

\subsubsection{The ANOVA-DDP and linear DDP models}\label{sect:ANOVADDP}

One of the earliest versions of DDP models was the ANOVA-DDP of
\cite{deiorio;mueller;rosner;maceachern;2004}. Let $\yb=(y_1,\ldots,y_n)$
be a vector of responses (possibly vector-valued) for each of $n$
subjects, and suppose that $\xb=(\xb_1,\ldots,\xb_n)$ is a corresponding
set of covariates. Assume each $\xb_i$ is in turn a vector of $c$
categorical covariates, $\xb_i=(x_{i1},\ldots,x_{ic})$. Interpret $\xb_i$
as factors in an ANOVA model, and let $d_i$ denote corresponding design
vectors. Assume then that $\xb_i$ contains all the desired main effects
and interactions, as well as desired identifiability constraints. Note
that the covariate space $\XX$ in this setup is discrete, and so we have a
finite number of RPMs. The idea of the ANOVA-DDP models is to encode the
covariate dependence in the form of simple linear regressions for the atom
processes $\{\theta_h(\xb):\,\xb\in\XX\}$. Specifically, this approach
uses $\theta_h(\xb)=\lambda_h ' d_{\xb}$ for $h\ge 1$ where
$\{\lambda_h:\,h\ge 1\}$ is a sequence of iid random vectors with
distribution $G_0$ and $d_{\xb}$ is the design vector that corresponds to
a generic combination of observed factorial covariates $\xb$. The model
just described implies that \eqref{eq:singleweight} becomes
$$G_{\xb}(\bullet)=\sum_{h=1}^{\infty}w_h \delta_{\lambda_h'
  d_{\xb}}(\bullet),
$$
i.e., a DP mixture of linear models $\lambda_h'd_{\xb}$. Each element of
the  
collection $\GG=\{G_{\xb}:\,\xb\in\XX\}$ has a DP prior distribution with
atoms given by $\{\lambda_h' d_{\xb}:\, h\ge 1\}$. The elements of $\GG$
are correlated because they share a common set of weights and the atoms
are originated as linear combinations computed from a single set of
parameters, namely $\{w_h:\, h\ge 1\}$ and $\{\lambda_h:\, h\ge 1\}$.

 To accommodate a continous response,  
\cite{deiorio;mueller;rosner;maceachern;2004}
extended the above construction  through a convolution with a
continuous kernel, e.g., a normal kernel, leading to 
$$
y_i \mid G_{\xb_i} \overset{ind.}{\sim}
\int N\left(y_i\mid {m \mu} ,\phi \right)\, dG_{\xb_i}({m \mu})=
\int N\left(y_i\mid \lambda' d_{\xb_i},\phi \right)\, dG(\lambda).
$$
 The model can be restated by breaking the mixture with the
introduction of latent parameters: 
\begin{equation}\label{eq:deiorio}
y_i\mid \lambda_i,\phi\sim N(\lambda_i' d_i,\phi),\quad \lambda_1,\ldots,\lambda_n\mid G\iid G,
\quad G\sim DP(M,G_0).
\end{equation}
 The last expression highlights the nature of the model as just a DP
mixture of, in this case, normal linear models. The same simplification is
possible whenever the  atoms
$\{\theta_h(\xb):\,\xb\in\XX\}$ are   indexed by a finite-dimensional
parameter vector, like the linear model $\theta_h(\xb)=\lambda_h'd_{\xb}$
in this case. The model in \eqref{eq:deiorio} is completed  with a
suitable prior for the precision parameter $\phi$, e.g. $\phi\sim Ga(a,b)$
if a scalar, or $\phi\sim {\rm Wishart}(\nu,S)$ if a matrix. The above
model can be easily modified to mix over scale parameters as well. An
immediate consequence of \eqref{eq:deiorio} is that the induced marginal
distribution for a single response $y$ with design vector $d_{\xb}$ then
becomes a flexible infinite mixture model:
\begin{equation}\label{eq:margresponse}
y\sim \sum_{h=1}^{\infty} w_h N(y\mid \lambda_h' d_{\xb},\phi).
\end{equation}
We remark here that the hierarchical structure leading to
\eqref{eq:deiorio} reflects a common practice in the use and application
of the DDP. Since marginally each element of the $\GG$ family is almost
surely discrete (because it is drawn from a DP), models for discrete
outcomes are frequently built on convolving the DPs with a continuous
kernel, thus yielding a mixture of continuous distributions, which is
itself a continuous distribution. In the ANOVA-DDP model of
\cite{deiorio;mueller;rosner;maceachern;2004}, the normal kernel plays
precisely this role.

\cite{DEQUMU:07} applied the ANOVA-DDP construction to model random
effects for longitudinal hormone profiles of pregnant women, where the
dependence was on a normal/abnormal pregnancy indicator. This setting was
particularly useful for classification purposes.   More recently,
\cite{gutierrez2019} use the ANOVA-DDP framework to propose a
multiple testing procedure for comparing several treatments against a control.
A further extension of the ANOVA-DDP construction was given in
\cite{deiorio;johnson;mueller;rosner;2009}, who  considered the modeling of
nonproportional hazards for survival analysis. They considered a cancer
clinical trial, where interest centered on whether high doses of a
treatment are more effective than lower doses. The data included
additional discrete and continuous covariates, so the model was under the
extended ANCOVA-style framework that adds linear combinations of
continuous covariates to the ANOVA factorial design.

This same idea can be extended to linear combinations of any given set of
covariates, giving rise to the {\em linear} DDP (LDDP). Specifically, such
models involve a linear combination of a set of covariates, as in, e.g.
general linear models, and so the infinite mixture on the right-hand side
of \eqref{eq:margresponse} becomes $\sum_{h=1}^{\infty} w_h
N(y\mid\lambda_h' \xb,\phi)$, where $\xb$ is now the generic value of the
(typically vector-valued) covariate.  As earlier, the weights $\{w_h\}$
follow a DP-style stick-breaking specification. An analogous expression
for a more general kernel function $k$  can be immediately
derived. The same type of construction was explored in
\cite{jara;lesaffre;deiorio;quintana;2010} in the context of doubly
censored outcomes. Their model involves an interval-valued response,
corresponding to the observed onset and event times (cavities in the teeth
of children from Flanders, Belgium, in their example). Associated with
each such response is a latent bivariate vector of true onset and event
times, and these are modeled (in the logarithmic scale) using a linear DDP
defined in terms of covariates that include deciduous second molars health
status and the age at which children started brushing.

\subsubsection{ Spatial DDP}
\label{sect:DDPGP}  \cite{gelfand;kottas;maceachern;2005} define what
can be interpreted as a spatial case of a common weight DDP
\eqref{eq:singleweight} for $G_{s}$, with $s \in D \subset \bsc{R}^d$
being spatial locations and $\th_h(s)$ generated by a baseline GP, as in
the common-weight DDP. However, the focus is not on $G_{s}$ as in
\eqref{eq:singleweight}, but instead on $\thb_D \sim \sum w_h
\delta_{\thb_{h,D}}$, where $\thb_{h,D} = \{\th_h(s),\; s \in D\}$. Let
$\sb=(s_1,\ldots,s_n)$ denote a set of $n$ locations at which observations
$\yb=(y_1,\ldots,y_n)$ are made. They consider repeat observations
$\yb_t$, $t=1,\ldots,T$, with occasion-specific covariates $\xb_t$.
Writing a mixture with respect to a DP random measure as a hierarchical
model, they assume 
$$
\yb_t \mid \thb_t, \betab, \tau^2 \ind N(\xb_t' \betab +\thb_t, \tau^2
\bm{I}), \quad \thb_t\mid G^\eta\iid G^\eta,\quad G^\eta\sim
DP(M,G_0^\eta),$$ where $G_0^\eta\equiv N(\bm{0},\sigma^2 \bm{H}(\eta))$
and $\bm{H}(\eta)$ is a suitable covariance function depending on
hyperparameters $\eta$.

\cite{dunson;herring;2006} considered a model for a collection of random
functions based on a finite set of latent trajectories described by
Gaussian processes. The observations are thus seen as arising from the
convolution of a smooth latent trajectory and a noisy Gaussian process.
Their motivation came from the study of the relationship between
disinfection by-products in the water in early pregnancy and later
outcomes. Specifically, denoting by $g_i$ the stochastic process, i.e.
$\{g_i(t):\,t>0\}$, associated with subject $1\le i\le n$,
\cite{dunson;herring;2006} assume that
$$g_i=\gamma_i+\epsilon_i,\quad \gamma_i\iid G,\quad \epsilon_i\iid GP(\bm{H}(\eta)),$$
where $\gamma_i$ is the latent trajectory, and $GP(\bm{H}(\eta))$ denotes
a Gaussian process with covariance function $\bm{H}(\eta)$. Their approach
specifies the RPM $G$ as $G(\cdot)=\sum_{h=1}^k p_h
\delta_{\Theta_h}(\cdot)$ with $\Theta_h\sim GP(\bm{H}(\eta_{\kappa_h}))$,
i.e., a finite mixture of atoms given by Gaussian processes with suitable
covariance functions. By choosing $\kappa_h=\kappa$ for all $h$ and
$(p_1,\ldots,p_k)\sim Dir(M/k,\ldots,M/k)$, the resulting RPM $G$
approaches $G(\cdot)=\sum_{h=1}^{\infty}w_h \delta_{\Theta_h}(\cdot)$ as
$k\rightarrow\infty$ with DP-style weights \citep[see,
e.g.][]{green-richardson:2001}.

\subsubsection{Dynamic DDP}\label{sect:DynDDP}

The DDP framework has also been used to model dynamic phenomena, by means
of a sequence of random distributions that evolve in time.
\cite{caronetal:08} considered a dynamic linear model formulation to solve
this problem, where the state and observation noise distributions where
modeled as DP mixtures using two independent DPs so that the mean of the
underlying processes is allowed to change in time.

\cite{rodriguez-terhorst:08} considered a related model, based on a DDP
formulation, where now the atoms in the infinite mixture are allowed to
change in time. Letting $y_{it}$ denote the $i$th observation at time
$1\le t\le T$, they proposed the model
$$y_{it}\mid G_t\sim \int N(\bm{F}_{it}' \thb_t,\sigma^2)\,dG_t(\thb_t,\sigma^2),\quad
G_t(\cdot)=\sum_{h=1}^{\infty} w_h
\delta_{(\thb^*_{ht},\sigma^{*2}_h)}(\cdot),\quad \thb^*_{ht}\sim
N(\bm{H}_t\thb^*_{h,t-1},\sigma^{*2}_h\bm{W}_t),$$ completed with
conjugate priors for $\sigma^{*2}_h$ and $\thb^*_{h,0}$. Matrices
$\bm{F}_{it}$, $\bm{H}_{t}$ and $\bm{W}_{t}$ are assumed known and can be
used to represent many patterns such as trends, periodicity, etc. The
resulting model for $\GG=\{G_t:\,1\le t\le T\}$ is thus a DDP, where the
components of the atoms controlling the distribution means evolve in time
in an autoregressive fashion.

\cite{diluccaetal:12} considered a model for a sequence of random
variables $\{y_t:\,t\ge 1\}$ featuring a general autoregressive
formulation by means of $y_t\mid (y_{t-1},\ldots,y_{t-p})=\bm{y}\sim
G_{\bm{y}}$ and the problem of defining a prior for
$\GG=\{G_{\bm{y}}:\,\bm{y}\in\YY\}$. They discussed a general prior DDP
model of the form
$G_{\bm{y}}(\cdot)=\sum_{h=1}^{\infty}w_h(\bm{y})\delta_{\bm{y}}(\cdot)$.
\cite{lau&so:08} considered similar types of model, where each atom can be
expressed as an infinite mixture of autoregressions of order $p$.
\cite{diluccaetal:12} focused on the particular single-weights case and an
order $p=1$ process where the atom processes are expressed as simple
linear autoregression: $\theta_h(\bm{y})=\beta_h+\alpha_h y$. The full
model in this case can be expressed as
\begin{equation}\label{eq:lucca}
y_t\mid y_{t-1}=y,\alpha_t,\beta_t,\sigma^2\sim N(\beta_t+\alpha_t y,\sigma^2),\quad
(\beta_t,\alpha_t)\mid G\iid G,\quad G\sim DP(M,G_0).
\end{equation}
However, they also
considered the case when atoms are defined as $\theta_h(y)=b+a_h
y+OU(\rho,\tau^2)$, where $OU(\rho,\tau^2)$ denotes the Ornstein-Uhlenbeck
process, a particular Gaussian process with covariance function of the
form $Cov[\theta(s),\theta(t)]=\tau^2 \rho^{|s-t|}$. \cite{diluccaetal:12}
extended this approach for sequences of binary outcomes defined in terms
of an autoregressive process $Z_t$ with a flexible DDP prior distribution,
where dependence is on the previous $p$ binary responses.

 An interesting variation of a dynamic DDP construction is proposed by
\cite{ascolani2020} who define a family $\GG=\{G_t,\; t\ge 0\}$ of
dependent random probability measures indexed by time. Their
construction is motivated by a Fleming-Viot process. The random
probability measures $G_t$ share some, but not all atoms. The set $D_t$ of
atoms in the original $G_0$ which are shared in $G_t$ is defined as a pure
death process over time. Importantly, each $G_t$ marginally remains a DP
random measure. They refer to the model as the Fleming-Viot-DDP. In
\cite{pruenster2013} this construction is applied to model market shares
over time. \cite{mena2016} construct another common-atoms DDP over time by
setting up a Wrights-Fisher diffusion on the fractions $v_{t,\ell}$ in the
stick-breaking construction of the marginal DP prior for $G_t$. 




\subsection{The single-atoms DDP}\label{sect:saDDP}

A parallel construction  to the common weights DDP in the previous
section  considers a set of common atoms across all values of
$\bm{x}$. This is the so called ``single-atoms'' DDP model, for which
\eqref{ddp_def}  takes  the form
\begin{eqnarray}\label{eq:singleatoms}
G_{\xb}(\bullet) = \sum_{h=1}^{\infty}  \underbrace{\left\{V_h(\xb)
\prod_{\ell < h} \left[1 - V_{\ell}(\xb) \right] \right\}}_{w_h(\xb)} \delta_{\btheta_h}(\bullet),
\end{eqnarray}
where $V_h(\xb)$, $h \in \mathbb{N}$, are $[0,1]$-valued independent
stochastic processes with index set $\XX$ and 
marginal distributions $\Be(1,M_{\xb})$.  The locations
$\btheta_h$, $h \in \mathbb{N}$, are independent
with marginal distributions  $G^0$;  and
the $\{V_h(\xb)\}$ and $\{\btheta_h\}$ collections are mutually
independent. 

Under the single-atoms model, all the covariate-dependence is expressed
through the weights of the stick-breaking representation. One advantage of
doing so is that, unlike the single-weights case,  the implied prior
probability model on partitions changes with the values of $x\in\XX$. This
is important when the implied partition is of interest. Another important
feature is that problems related to extrapolation of $\btheta_h(\xb)$ are
avoided, which could otherwise arise for inference for a new value of
$\xb$ beyond the range of the observed data. This is the case because
under the single-atoms DDP all atoms are linked with observed data, in
contrast to the single-weights DDP which includes atoms for new covariate
values that are not linked with any observed data.

\cite{duan;guidani;gelfand;2007} describe a model motivated by the
analysis of spatially varying responses. Let $\{y(s):\,s\in D\}$ be a
stochastic process indexed by locations in a set $D\subset\bsc{R}^d$, and
let $s_1,\ldots,s_n$ the locations at which observations are collected.
Their general construction involves a RPM $G$ over the space of surfaces
of $D$ having finite-dimensionals adopting the following form: for any
$s_1,\ldots,s_n\in D$ and $A_1,\ldots,A_n$ Borel-measurable sets in
$\bsc{R}$,
$$P(y(s_1)\in A_1,\ldots,y(s_n)\in A_n)=\sum_{i_1=1}^{\infty}\cdots \sum_{i_n=1}^{\infty}
p_{i(s_1),\ldots,i(s_n)} \delta_{\theta_{i(s_1)}}(s_1)\cdots
\delta_{\theta_{i(s_n)}}(s_n),$$ where the $\theta_j$'s are iid from $G_0$
and the weights $\{p_{i(s_1),\ldots,i(s_n)}\}$ determine the site-specific
joint selection probabilities. Conditions can be given so that the above
specification follows a DP at any given location.

Always in the spatial context, specifically of modeling for hurricane
surface wind fields, \cite{reich&fuentes:07} propose a general framework
that includes the single-atoms DDP as a special case. Their model is
specially designed for spatial dependence as well, so that the covariates
are geographical coordinates. Letting $\sb$ denote such coordinates, their
construction involves weights computed as $w_1(\sb)=V_1(\sb)$ and
$w_h(\sb)=V_h(\sb)\prod_{\ell=1}^{h-1}(1-V_{\ell}(\sb))$ for $h>1$, where
$V_h(\sb)=\omega_h(\sb)V_h$, and $V_h\iid {\rm Beta}(a,b)$. The function
$\omega_h(\sb)$ is centered at knot $\bm{\psi}_h=(\psi_{h1},\psi_{h2})$,
and the spread is controlled by parameters $\bm{e}_h=(e_{h1},e_{h2})$.
\cite{reich&fuentes:07} discuss several possible choices for the
$\omega_h$ functions and related parameters.


\cite{griffin;steel;2006} define another interesting variation of the
basic DDP by keeping both sets of parameters, locations and  the
fractions ($V_h$),  unchanged across $\bm{x}$. They use instead
permutations of how the weights are matched with locations. The
permutations change with $\bm{x}$. One advantage of such models is the
fact that the support of $G_{\bm{x}}$ remains constant over $\bm{x}$, a
feature that can be important for extrapolation beyond the observed data.
A modification of this idea was explored by \cite{griffin&steel:10} to
generate what they called the DP regression smoother.  The construction is
centered over a class of regression models, and dependence is on the
weights. More recently, similar ideas are used by \cite{griffin&steel:11}
to construct a family of prior distributions for a sequence of time
dependent general RPMs that include the DDP setting as a special case.
Another simple sequence of time-dependent DDPs was proposed by
\cite{gutierrez2016time}, with a Markov chain structure for the sequence
of time-varying sticks, and with application to the analysis of air
quality data.

\section{Variations of MacEachern's DDP} \label{variationDDP}

In this section we discuss a variety of models extending the original
definition \eqref{ddp_def}. Many of these extensions are based on
constructing independent weights and atoms processes indexed by
covariates, but that do not necessarily produce a DP-distributed random
measure. From an intuitive viewpoint, these classes of models can be seen
as taking the basic DP construction and altering some of their basic
components in terms of predictors $\xb\in\XX$ to a form that may differ
from the initial distributional properties. While this typically modifies
the marginal DP property, the extra flexibility allows one to tailor the
properties of the model to fit specific applications.

\subsection{Weighted mixture of DPs (WMDP)}

\cite{dunson;pillai;park;2007} proposed a data-based prior using the
observed predictors $\xb_1,\ldots,\xb_n$. For every $\xb \in \XX \subset
\mathbb{R}^p$, they considered the following construction
$$
G_{\xb}(\bullet) =\sum_{j=1}^n \left( \frac{\gamma_j K(\xb,\xb_j)}{\sum_{\ell=1}^n
\gamma_\ell K(\xb,\xb_\ell)}\right) G_j(\bullet),
$$
with
$$
\gamma_j \mid \kappa \overset{iid}{\sim}\Gamma(\kappa,n\kappa),\quad
G_j \mid M, G_0 \overset{iid}{\sim}DP(M, G_0),
$$
where $K: \XX \times \XX \longrightarrow \mathbb{R}^+$ is a bounded kernel
function. The choice of $K$ impacts the degree of borrowing of information
from the neighbors in estimating the distribution at any particular
predictor value $\xb$. Some choices are discussed in the original
technical report. In the paper, they considered
\begin{equation}\label{eq:Kdunson}
K\left(\xb,\xb'\right)=\exp \left\{\psi ||  \xb-\xb'||^2 \right\},\quad
\psi \mid \mu_{\psi},\sigma^2_{\psi} \sim LN(\mu_{\psi},\sigma^2_{\psi}),
\end{equation}
where $LN(a,b)$ denotes the log-normal distribution with parameters
$a\in\RR$ and $b>0$. With this choice, the resulting model for a given
$\xb$ borrows more heavily from those $G_j$'s for which the corresponding
$\xb_j$ is close to $\xb$. One primary application of this particular
construction is in the context of {\em density regression} i.e. in
measuring how a probability distribution on the space of responses $\YY$
changes according to predictors $\bm{x}\in\XX$.

\subsection{Kernel stick-breaking}

The kernel stick-breaking process (KSBP) was introduced by
\cite{dunson;park;2008}. For all $\xb \in \XX \subset \mathbb{R}^p$, the
KSBP is defined as follows
\begin{equation}\label{eq:KSBP}
G_{\xb}(\bullet) =\sum_{h=1}^{\infty}\left\{ W(\xb; \boldsymbol{\Gamma}_h,V_h)
\prod_{\ell<h}\biggl(1- W(\xb; \Gamma_{\ell},V_{\ell})\biggr) \right\} G_h(\bullet),
\end{equation}
where $W(\xb; \Gamma_h,V_h)=V_h K(\xb,\boldsymbol{\Gamma}_h)$,  with $K:
\XX \times \XX \longrightarrow [0,1]$, e.g. as given in
\eqref{eq:Kdunson}, $V_h \mid a_h,b_h \overset{ind.}{\sim}\Be(a_h,b_h)$,
$\boldsymbol{\Gamma}_h \mid H\overset{iid}{\sim}H$ (random kernel
locations), and $G_h \mid \mathcal{G}\overset{iid}{\sim} \mathcal{G}$
(random probability measures). The KSBP thus begins with an infinite
sequence of basis random distributions $\{G_h\}$ and then constructs
covariate-dependent random measures by mixing according to distance from
the random locations $\Gamma_h$, with stick-breaking probabilities that
are defined as a kernel multiplied by Beta-distributed weights. It is also
possible to simplify the definition of KSBP, adopting the particular form
$$
G_{\xb}(\bullet) =\sum_{h=1}^{\infty}\left\{ W(\xb; \boldsymbol{\Gamma}_h,V_h)
\prod_{\ell<h}\biggl((1- W(\xb; \Gamma_{\ell},V_{\ell})\biggr) \right\} \delta_{\thb_h}(\bullet),
$$
where $W(\xb; \Gamma_h,V_h)=V_h K(\xb,\boldsymbol{\Gamma}_h)$,  with $K:
\XX \times \XX \longrightarrow [0,1]$, $V_h \mid M
\overset{iid}{\sim}\Be(1,M)$, $\boldsymbol{\Gamma}_h \mid
H\overset{iid}{\sim}H$ (random kernel locations), and $\thb_h \mid
G_0\overset{iid}{\sim}G_0$. This amounts to replacing the random measure
$G_h(\bullet)$ defined in \eqref{eq:KSBP} by just a single atom $\thb_h$.
Compared to the former, this latter version of KSBP greatly reduces model
complexity while still retaining some flexibility.

\subsection{Probit and logit stick-breaking}

\cite{chung;dunson;2009} introduced a modification of the stick-breaking
representation for DPs where the Beta random variables are replaced by
normally distributed random variables transformed using the standard
normal CDF. They refer to the resulting measure as the probit-stick
breaking (PSB) process. The PSB is defined by
\begin{equation}\label{eq:PSBP}
G(\bullet) = \sum_{h=1}^{\infty} \left\{\Phi(\eta_h) \prod_{\ell<h}(1-\Phi(\eta_{\ell}))\right\} \delta_{\thb_h}(\bullet),
\end{equation}
where $\eta_h \mid \mu \overset{iid}{\sim} N(\mu,1)$ and $\thb_h\mid G_0
\overset{iid}{\sim} G_0$. If $\mu=0$, \eqref{eq:PSBP} reduces to a regular
DP with $M=1$, i.e. uniformly distributed sticks. \cite{chung;dunson;2009}
also consider a covariate-dependent version of the PSB to model sets of
related probability distributions. This is done by replacing the $\eta_h$
variables with suitable stochastic processes or regression functions. For
instance, if $\{\eta_h(\xb): \xb \in \XX\}$ denote independent Gaussian
processes with unit variance, a dependent PSB can be defined as
\begin{equation}\label{eq:dPSBP}
G_{\xb}(\bullet) = \sum_{h=1}^{\infty} \left\{\Phi(\eta_h(\xb)) \prod_{\ell<h}\left[1-\Phi(\eta_{\ell}(\xb))\right]\right\}
\delta_{\thb_h}(\bullet).
\end{equation}
A similar modification can be obtained by taking $\eta_h(\xb)=\xb^T
\boldsymbol{\gamma}_h$. More generally, let
$\eta_h(\xb)=\alpha_h+f_h(\xb)$ with $\alpha_h\sim N(\mu,1)$ and
$f_h:\bsc{R}^p\rightarrow\bsc{R}$ an unknown regression function,
characterized by finitely many parameters $\bphi_h$, with
$\bphi_h\sim\bm{H}$. Denote this model as $\mbox{PSBP}(\mu,\bm{H},G_0)$.
One main focus of the proposal in \cite{chung;dunson;2009} was variable
selection. To that end, they assume the model
$$y\mid\xb\sim f(y\mid\xb)=\int N(y\mid \xb'\bbeta,\tau^{-1})\,dP_{\XX}(\bbeta,\tau),\quad
P_{\XX}=\{P_{\xb}:\,\xb\in\XX\}\sim \mbox{PSBP}(\mu,\bm{H},G_0),$$ where
the variable selection structure is here introduced in $\bm{H}$ and in
$G_0$, and by considering inclusion/exclusion indicators at the level of
the atoms in \eqref{eq:dPSBP}. See further discussion on PSBP in
\cite{rodriguez;dunson;2011}. A related construction, termed the
logit-stick breaking process was proposed in
\cite{ren;du;carin;dunson;2011}, which essentially replaces the probit by
a logit link in \eqref{eq:PSBP}. Applications of logit-stick breaking
processes to density regression can be found in \cite{RIGON2020}.

\subsection{Hierarchical mixture of DP}

Consider again the case $\XX=\{1,\ldots,J\}$, as in the example presented
in Section~\ref{sect:intro}, and let $\GG=\{G_{\xb} : \xb \in
\XX\}=\{G_1,\ldots,G_J\}$. Motivated by the need to borrow strength across
related studies (a situation also arising in applications of
meta-analysis), \cite{mueller;quintana;rosner;2004} proposed a
hierarchical DP model. In this construction, the probability distribution
for group $j$ is a weighted mixture of independent random measures.
Specifically, the probability model for a group is defined as a mixture of
a common distribution $H_0$, shared by all groups, and an idiosyncratic
component $H_j$, which is specific to each group,
\begin{equation}\label{eq:epsilon}
G_j(\bullet) =\epsilon H_0(\bullet)+ (1-\epsilon) H_j(\bullet),
\end{equation}
where $\epsilon \in [0,1]$ controls the level of dependence in the set
$\GG$, and $H_0$, $H_1,\ldots, H_J$ are assumed to be independent DPs. The
two extreme cases depicted in Figure~\ref{fig:GorGj} correspond to
$\epsilon=1$ for panel (a), i.e. a single common measure, and $\epsilon=0$
for panel (b), i.e. independent model and no borrowing of strength.
Model~\eqref{eq:epsilon} represents then a trade-off between these two
extreme options, allowing one to borrow strength through the common part,
while retaining flexibility for the study-specific part of the model. More
recently,~\cite{wang&rosner:19} used this construction to propose a
propensity score-based mixture model to combine subject-level information
from randomized and registry studies, their goal being inference on a
causal treatment effect.

Extending \eqref{eq:epsilon} to the case of continuous predictors can be
easily accomplished by combining a study index, $j$, continuous predictors
$\zb$, and setting up
$$
G_{j,\zb}(\bullet) =\epsilon H_{0,\zb}(\bullet)+ (1-\epsilon) H_{j,\zb}(\bullet),
$$
where  $H_{0,\zb}$, $H_{1,\zb},\ldots, H_{J,\zb}$ are now independent
MacEachern's DDPs based on the continuous predictors $\zb$, incorporating
dependence on predictors as in the LDDP or ANCOVA-DDP of
Section~\ref{sect:ANOVADDP}, according to the available covariates types.
The construction is easily modified to allow for study-specific variation
in the weight assigned to the idiosyncratic component $H_j$ by replacing
$\epsilon$ with $\epsilon_j$.

A clever variation of this construction is introduced in
\cite{kolossiatis&al:13} who chose the weight $\eps$ to ensure that $G_j$
remains marginally a DP again. A more general version of the same
construction appears in \cite{camerlenghi2019}.

\subsection{Hierarchical DP of \cite{teh;jordan;beal;blei;2006}}

In the context of $\XX=\{1,\ldots,J\}$, \cite{teh;jordan;beal;blei;2006}
proposed a model that induces an ANOVA type of dependence. In their
construction, referred to as the hierarchical DP (HDP), the random
probability measure for the $j$th group $G_j$, $j=1,\ldots,J$, is a DP
conditional on a common measure $G$, which in turn is also a DP,
\begin{equation}\label{eq:HDP}
G_j \mid M_j, G \overset{ind.}{\sim} DP(M_j, G),\qquad j=1,\ldots,J,\qquad
 G\mid M, G_0 \sim DP(M, G_0).
\end{equation}
A main motivation behind the particular form adopted in~\eqref{eq:HDP} was
to provide a model that allows for sharing clusters among related
subpopulations. \cite{teh;jordan;beal;blei;2006} consider the analysis of
text, where a primary goal was to share clusters among various documents
within a cluster, and also to share clusters among various corpora. The
HDP facilitates the construction of clusters at various levels, due to its
hierarchical formulation. In fact, this clustering structure can be
described in terms of a Chinese restaurant franchise, where at each of a
collection of restaurants customers sit at tables organized by dishes, and
dishes can be ordered from a global menu available to all restaurants.
This construction, if restricted to a single restaurant, reduces to the
usual Chinese restaurant process~\citep{aldous:85} that is colloquially
used to describe the DP.

\subsection{The nested DP}

Also in the context of $\XX=\{1,\ldots,J\}$,
\cite{rodriguez;dunson;gelfand;2008} proposed an alternative model,
referred to as the nested DP.  In their construction the law of the random
probability measure for the $j$th group $G_j$, $j=1,\ldots,J$, is an
infinite mixture of trajectories of DPs,
\begin{equation}\label{eq:NDP}
G_j  \overset{ind.}{\sim} \sum_{h=1}^{\infty} \pi_h \delta_{G_h^*} (\bullet),\qquad j=1,\ldots,J,
\qquad G_h^* \mid M_2, H \overset{i.i.d.}{\sim} DP(M_2, H),
\end{equation}
where $\pi_h= V_h \prod_{\ell<h} (1-V_\ell)$, with $V_h \mid M_1
\overset{i.i.d.}{\sim}\Be(1,M_1)$, for $h=1,2,\ldots$. The main motivation
behind \eqref{eq:NDP} was to construct a clustering of individuals across
the different groups, e.g. patients within different medical centers. The
NDP model aims to simultaneously cluster patients within centers,
borrowing information across centers for which similar clusters are
detected, and to cluster different centers. This is then a type of
multilevel clustering.

By way of comparison, it can be noted that in the HDP of
\cite{teh;jordan;beal;blei;2006}, the random measures in
$\GG=\{G_1,\ldots,G_J\}$ share the same atoms but assign them different
weights, while in the NDP two distributions $G_{j_1}$ and $G_{j_2}$ either
share both atoms and weights (i.e. they are identical) or share nothing at
all. Thus, the NDP allows for clusters at the level of the responses and
also at the level of distributions, while the HDP allows for clusters only
at the level of observations.

One of the limitations of the NDP is that for any two random measures
$G_{j_1}, G_{j_2}$ it supports only the two extreme cases of either all
atoms and weights shared, i.e., $G_{j_1}=G_{j_2}$, or no atoms shared, but
does not allow any intermediate configuration with some atoms being
shared. As a consequence,
whenever there are ties of atoms between $G_{j_1}$ and $G_{j_2}$, the
nested structure forces the two random distributions to be identical. For
a discussion of this problem see \cite{camerlenghi2019} who introduce the
latent nested process as a more general hierarchical prior for random
probability measures that avoids this restriction.  More recently,
\cite{beraha2020semihierarchical} propose the semi-hierarchical DP as an
alternative solution to the limitations inherent to latent nested
processes, with the added benefit of computationally efficient
implementations to the comparison and clustering of potentially many
subpopulations. 

Like any discrete random probability measure, the NDP can be used to
define random partitions. Model \eqref{eq:NDP} could be written in short
as $G_j \sim \DP\{ M_1, \DP(M_2, H) \}$. The outer DP, with total mass
$M_1$ gives rise to a partition of $\XX$. Consider now samples $y_{ji}
\sim G_j$, $i=1,\ldots,n_j$. The inner DP gives rise to random partitions
of $\YY_j=\{1,\ldots,n_i\}$, i.e., the NDP defines a nested partition of
$\XX$ and $\YY_j$, with the prior for the random partitions for $\YY_j$
and $\YY_{j'}$ being equal in distribution when $G_j=G_{j'}$. Curiously,
exactly the same random nested partition on $\XX$ and $\YY_j$ is implied
by the enriched DP (EDP) defined in \cite{wade2011}. The EDP defines a
random probability measure for pairs $(x_i,y_i)$ as $P_X(x_i)\,
P_{Y|X}(y_i \mid x_i)$, which, as discrete random probability measures,
gives rise to the same random nested partition.

\subsection{The product of independent DPs}

Alternatively, \cite{gelfand;kottas;2001} proposed an approach based on
the product of independent random measures. In this construction the
distribution for the $j$th group $G_j$, $j=1,\ldots,J$, is given by
$$
G_j(\bullet) \equiv H_j (\bullet)\prod_{\ell< j} H_{\ell} (\bullet),\ \ \ j=1,\ldots,J,
$$
where
$$H_j \mid M_j , H_{0j} \overset{ind.}{\sim} DP(M_j, H_{0j}), \ \ \ j=1,\ldots,J.$$
The motivation for this construction arises from the need to define models
that induce stochastic ordering for the random group specific
distributions $G_j$. The ordering holds with probability $1$ in the prior
and so is also satisfied a posteriori.

\subsection{Other constructions}\label{sect:other}

\cite{chung;dunson;2009} proposed a similar construction, referred to as
the local DP, where the stick-breaking weights selected to define the
probability weights depend on a set of random locations and their
distances to a given predicted value. In this construction, the support
points also depend on predictors.

\cite{fuentes;mena;walker;2009} considered a dependent variation of
geometric-weights stick-breaking processes
\citep{mena;ruggiero;walker;2011}. In this construction, the
stick-breaking weights are replaced by their expected value, thus reducing
the number of parameters.

Dependent neutral to the right processes and correlated  two-parameter
Poisson-Dirichlet processes have been proposed by
\cite{epifani;lijoi;2010} and \cite{leisen;lijoi;2011}, respectively, by
considering suitable L\'evy copulas. The general class of dependent
normalized completely random measures has been discussed, for instance, by
\cite{lijoi;nipoti;pruenster;2014}.

Another type of construction stems from the fact that
the Dirichlet process is also a special case of a
normalized random measure with independent increments (NRMI), as described
in~\cite{regazzini;etal;2003}. This means that if $F$ has a DP
distribution, then it can be expressed in the form
$$F(\bullet)=\frac{\mu(\bullet)}{\mu(\Omega)},$$
where $\Omega$ is the space where the DP is defined, and $\mu$ is a
completely random measure on $(\Omega,\mathcal{B}(\Omega))$, that is, for
any collection of disjoint sets $A_1,A_2,\ldots$ in $\mathcal{B}(\Omega)$,
the Borel $\sigma$-field in $\Omega$, the random variables
$\mu(A_1),\mu(A_2),\ldots$ are independent, and $\mu(\cup_{j=1}^{\infty}
A_j)=\sum_{j=1}^{\infty}\mu(A_j)$ holds true a.s. See,
e.g.~\cite{james-lijoi-pruenster:09}. As shown in \cite{ferguson;73}, the
Dirichlet process arises as the normalized version of a Gamma process.
 \cite{barriosNRMI2013}, \cite{favaro2013} and
\cite{Argiento:2010}  
discuss modeling with mixtures of NRMIs, and
in particular discuss practical implementation of posterior simulation for
such models. See additional MCMC implementation details in~
Building on related ideas,  \cite{epifani;lijoi;2010} and
\cite{leisen;lijoi;2011}, 
proposed  dependent neutral to the right processes and correlated
two-parameter Poisson-Dirichlet processes,  respectively, by considering
suitable L\'evy copulas. A more general class of dependent normalized
completely random measures has been discussed, for instance, by
\cite{lijoi;nipoti;pruenster;2014}. This construction has also motivated
work on defining DDPs by way of introducing dependence in NRMIs.
\cite{NIPS2010_4151} used this idea to propose a Markov chain of Dirichlet
processes, and other extensions to normalized random measured are
described in \cite{ICML2012Chen_476} and in~\cite{pmlr-v28-chen13i}.

\section{The induced conditional density approach}\label{joint}

The approaches described so far yield valid inferences when the set of
predictors $\xb$ are fixed by design or are random but exogenous. Notice
that the exogeneity assumption permits us to focus on the problem of
conditional density estimation, regardless of the data generating
mechanism of the predictors, that is, if they are randomly generated or
fixed by design \citep[see, e.g.,][]{barndorff;1973,barndorff;1978}. Under
the presence of endogenous predictors, both the response and the
predictors should be modeled jointly.

In the context of continuous responses and predictors,
\cite{mueller;erkanli;west;1996} proposed a DPM of multivariate Gaussian
distributions for the complete data $\boldsymbol{d}_i=(y_i,\xb_i)'$,
$i=1,\ldots,n$, and looked at the induced conditional distributions.
Although \cite{mueller;erkanli;west;1996} focused on the mean function
only, $m(\xb) = E(y \mid \xb)$, their method can be easily extended to
provide inferences for the conditional density at covariate level $\xb$.
The model is given by
\begin{eqnarray}\nonumber
\boldsymbol{d}_i \mid G
\overset{iid}{\sim}
\int N_k \left(\boldsymbol{d}_i\mid \mub,\Sigb \right)  d G(\mub,\Sigb),
\end{eqnarray}
and
\begin{eqnarray}\nonumber
G \mid M, G_0 \sim DP\left(M, G_0 \right),
\end{eqnarray}
where $k=p+1$ is the dimension of the complete data vector
$\boldsymbol{d}_i$, and the baseline distribution $G_0$ is the conjugate
normal-inverted-Wishart (IW) distribution $ G_0 \equiv N_k\left(\mub \mid
\boldsymbol{m}_1, \kappa_0^{-1}\Sigb\right) \times IW_k\left(\Sigb \mid
\nu_1,\boldsymbol{\Psi}_1\right)$. The model is completed with
conditionally conjugate priors and hyperpriors on $m_1$, $\kappa_0$ and
$\Psib$, and, if desired, a gamma hyperprior on $M$.
The model induces a weight-dependent mixture model for the
regression,
\begin{equation}\label{eq:conditional}
  f_{\xb} (y)=\sum_{h=1}^\infty \omega_h(\xb) N(y \mid \beta_{0h} + \xb' \bfbeta_h,\sigma^2_h), 
\end{equation}
where
\begin{eqnarray}\nonumber
\omega_h(\xb)=\frac{w_h N_p(\xb\mid \mub_{2h} ,\Sigb_{22h})}
{\sum_{\ell=1}^\infty w_{\ell} N_p(\xb \mid \mub_{2\ell} ,\Sigb_{22\ell})},\qquad h=1,2,\ldots,
\end{eqnarray}
$\beta_{0h}=\mu_{1h}-\Sigb_{12h}\Sigb_{22h}^{-1}\mub_{2h}$,
$\boldsymbol{\beta}_{h}=\Sigb_{12h}\Sigb_{22h}^{-1}$, and
$\sigma^2_{h}=\sigma^2_{11h}-\Sigb_{12h}\Sigb_{22h}^{-1}\Sigb_{21h}$.
Here, the weights $w_h$ follow the usual DP stick-breaking construction,
and the remaining elements arise from the standard partition of the
vectors of means and (co)variance matrices given by
\begin{eqnarray}\nonumber
\mub_h =\left(
\begin{array}{c}
  \mu_{1h} \\
  \mub_{2h}
\end{array}
\right) \mbox{\ \ and \ \ }
\Sigb_h =
\left(
\begin{array}{cc}
  \sigma^2_{11h} & \Sigb_{12h} \\
  \Sigb_{21h} & \Sigb_{22h}
\end{array}
\right),
\end{eqnarray}
respectively.

The induced conditional density approach of
\cite{mueller;erkanli;west;1996} can be easily extended to handle mixed
continuous, $\xb_C$, and discrete predictors, $\xb_D$, by considering a
DPM model of product of appropriate kernels for discrete $k_D$ and
continuous $k_D$ variables,
\begin{eqnarray}\label{eq:condjoint}
\boldsymbol{d}_i \mid G
\overset{iid}{\sim}
\int k_D(\xb_{iD}\mid \thb_1)  k_C(y_i,\xb_C\mid \thb_2)\,d G(\thb_1,\thb_2),
\end{eqnarray}
i.e., assuming a multiplicative structure in the joint model for
$(y,\xb_D,\xb_C)$ that mimics conditional independence of $(y,\xb_C)$ and
$\xb_D$ given suitable parameter vectors $\thb_1$ and $\thb_2$.
Similar types of models, but looking only at the induced partition
structures, are discussed in \cite{mueller;quintana;2010}. In particular,
\cite{mueller;quintana;rosner;2011} proposed a version of
\eqref{eq:condjoint} that may be viewed as integrating out the random
measure $G$ in \eqref{eq:condjoint}, retaining only the random partition
model, while still allowing for covariate dependence in the prior. This
approach exploits the connection between the DP and product partition
models. See, e.g., \cite{quintana;iglesias;2003}.

We introduced the conditional density regression approach assuming
endogenous predictors, when the construction of a joint probability model
for $(y_i, \xb_i)$ is natural. However, the same construction can be used
to achieve the desired smooth locally weighted mixture of linear
regressions even when the $\xb_i$ are exogenous, or even if they are not
random at all. The choice of model depends largely on properties of the
model and ease of prior specification, tempered by computational concerns.

\section{Implied Random Partitions
  and other uses of the DDP model}
 \label{sec:part} One of the common applications of the DP mixture
model \eqref{eq:DPM} is to define a random partition and allow statistical
inference on such partitions. Consider an equivalent statement of i.i.d.
sampling from \eqref{eq:DPM} as a hierarchical model
\begin{equation}
  y_i \mid \th_i \sim p(y_i \mid \th_i)\quad
  \mbox{ and }\quad
  \th_i \sim G,
  \label{eq:DPMh}
\end{equation}
$i=1,\ldots,n$. The discrete nature of the DP random measure $G$ implies
positive probabilities of ties among the $\th_i$ with $K \leq N$ unique
values $\{\ths_1,\ldots,\ths_K\}$. Defining $S_j = \{i:\; \th_i=\ths_j\}$
defines a partition $\{1,\ldots,n\} = \bigcupdot_j S_j$. A common
application of the DP mixture model is to derive inference on such
partitions $\rho=\{S_1,\ldots,S_K\}$, and interpret the partitioning
subsets as meaningful subpopulations of the experimental units (e.g.,
patient subpopulations). In anticipation of the upcoming generalization to
the DDP, we introduce a slightly different but equivalent definition of
the clusters $S_j$. Recall the representation \eqref{eq:stbrDP} of  DP
random measure, $G = \sum w_h \delta_{\tht_h}$
Then the non-empty sets $R_h = \{i: \; \th_i=\tht_h\}$ describe the same
partition $\rho$. We switched from indexing clusters by their common
unique $\th_i$ values to identifying clusters by the matching atoms in
$G$. Similarly we can set up a model for independent sampling using a DDP
prior.  Specifically, consider 
\begin{equation}
  y_i \mid \th_i \sim p(y_i \mid \th_i)\quad
  \mbox{ and }\quad
  \th_i \mid x_i=x \sim G_x,
  \label{eq:DDPMh}
\end{equation}
$i=1,\ldots, n$, with a DDP prior on $\GG=\{G_x, x \in X\}$. For the
moment assume a categorical covariate $x_i \in \{1,\ldots, n_x\}$, and let
$G_x$, $x=1,\ldots,n_x$ denote the (marginal) random measures, and let
$I_x=\{i:\; x_i=x\}$ denote the subpopulation with covariate $x$. First,
by the earlier argument the model implies a random partition $\rho_x$ of
$I_x$, marginally, for each $x$. Indexing clusters by the corresponding
atom in $G_x$ implicitly defines a joint prior on $\{\rho_x,\; x \in X\}$,
or, alternatively, defines a partition of $\{1,\ldots,n\}$ with clusters
$S_j$ that cut across $I_x$. In particular, the model implies a joint
prior on $(\rho_x, \rho_{x'})$ for any $x \ne x'$, and it allows for
shared clusters across subpopulations.  Different assumptions
on various model aspects, such as dispersion in the baseline distribution,
or total mass parameter, would have a practical effect on the this
joint prior. Curiously, in contrast to 
the DP mixture model, the DDP model is not commonly used for inference on
these implied random partition(s).

Another feature of the DDP model is inference about distributional
homogeneity. To be specific, consider again the context of independent
sampling in \eqref{eq:DPMh} with a categorical covariate $x \in
\{1,\ldots, n_x\}$ and let $f_x(y) = \int p(y \mid \th) dG_x(\th)$ denote
the implied marginal distribution of $y_i \mid x_i=x$. In many
applications investigators might be interested in the event $f_x = f_{x'}$
for $x \ne x'$. While the DDP prior, short of a pathological special case,
implies zero prior probability for exact equality, posterior inference
includes meaningful posterior probabilities for $\{d(f_x,f_{x'})>\eps\}$
for any well defined distance of the two distributions. Specifics would
depend on particular applications. Related summaries, for example, by
displaying posterior means for $f_x$ over $x$ are shown in some papers
using DDP priors for density regression. See, e.g. \cite{gutierrez2019}.

\section{Application to Autoregressive Models} \label{applications}

We illustrate some of the discussed DDP-based nonparametric regression
models. We implement inference under the ANOVA-DDP or LDDP
model of \eqref{eq:deiorio} and
conditional density regression as in
\eqref{eq:conditional} to model (auto-)regression
on $x_t=y_{t-1}$ in time series data, using the LDDP model 
\begin{equation}\label{eq:AR1LDDP}
y_t\mid y_{t-1}=y,\beta_{t0},\beta_{t1},\sigma_t^2\sim N(\beta_{t0}+\beta_{t1} y,\sigma_t^2),\quad
(\beta_{t0},\beta_{t1},\sigma_t^2)\mid G\iid G,\quad G\sim DP(M,G_0(\cdot\mid\etab)),
\end{equation}
where $t=2,\ldots,n$, i.e. we mix over the  linear coefficients
and the variance.   The dependence in
\eqref{eq:AR1LDDP} is conveyed through linear functions of the first
lagged response in the atoms, keeping common weights. Here,
$G_0(\cdot\mid\etab)$ is the centering measure with hyperparameters
$\etab$.
 Following  
\cite{jara;hanson;quintana;mueller;rosner;2011} we use   $G_0 \equiv
N_2\left( \boldsymbol{\beta} \mid \mub_b, \boldsymbol{S}_b \right)
\Gamma\left(\sigma^{-2}\mid \tau_1/2,\tau_2/2\right)$, and complete
the prior specification as
\begin{eqnarray*}
M \mid a_0, b_0 \sim \Gamma\left(a_0,b_0\right), \quad &\, &
\tau_2 \mid \tau_{s_1}, \tau_{s_2} \sim \Gamma(\tau_{s_1}/2,\tau_{s_2}/2), \\
\mub_b\mid \boldsymbol{m}_0,\boldsymbol{S}_0 \sim N_p(\boldsymbol{m}_0,\boldsymbol{S}_0),\quad
&\, & \boldsymbol{S}_b\mid \nu,\boldsymbol{\Psi}\sim IW_p(\nu,\boldsymbol{\Psi}).
\end{eqnarray*}

For this illustration, we consider two specific datasets:

 Data set {\bf D1} are  the Old Faithful geyser data
\citep{hardle:91}, available 
as part of the \texttt{datasets} library available in \texttt{R},
consisting of $n=272$ observations on eruption times (in minutes) and
waiting times to the next eruption (also in minutes).
 Data set {\bf D2} is a time series of the Standard \& Poor's 500
index, from February 9, 1993 through February 9, 2015.
It is available in the \texttt{R} package {\em pdfetch}
\citep{pdfetch}, using the command
\begin{verbatim}
pdfetch_YAHOO("SPY",fields = "adjclose",
              from = as.Date("1993-02-09"), to = as.Date("2015-02-09"))
\end{verbatim}
In the following results we compare inference under model
\eqref{eq:AR1LDDP} with inference under density regression, as in
\eqref{eq:conditional}, again using $x_t=y_{t-1}$.
Recall that a conditional density approach 
is based on a DPM model for $\{(y_t,y_{t-1}):\,t=2,\ldots,n\}$. 

\begin{figure}
\centering
\subfigure[] 
{
    \includegraphics[width=5.5cm]{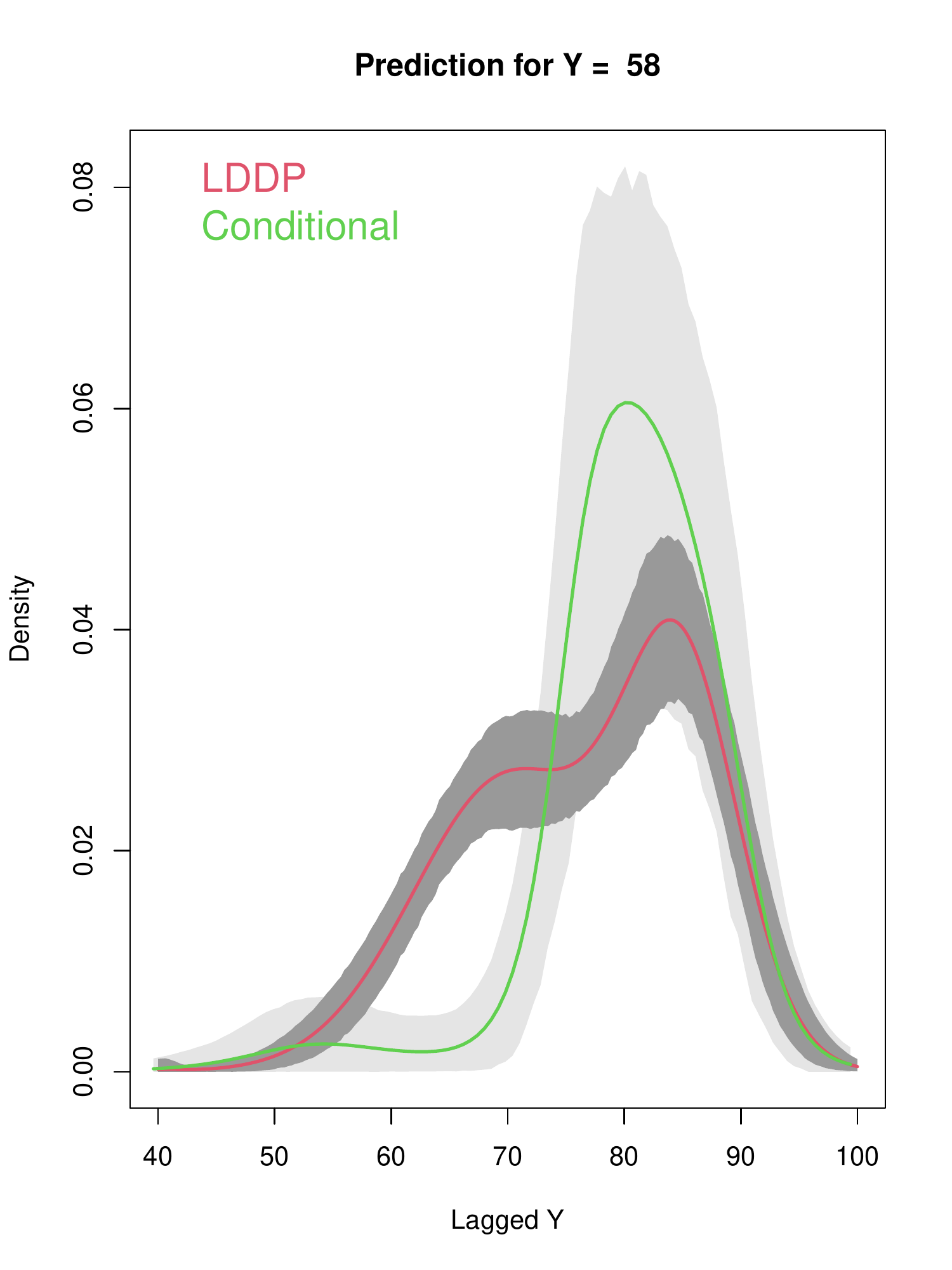}
}
\hspace{-0.5cm}\subfigure[] 
{
    \includegraphics[width=5.5cm]{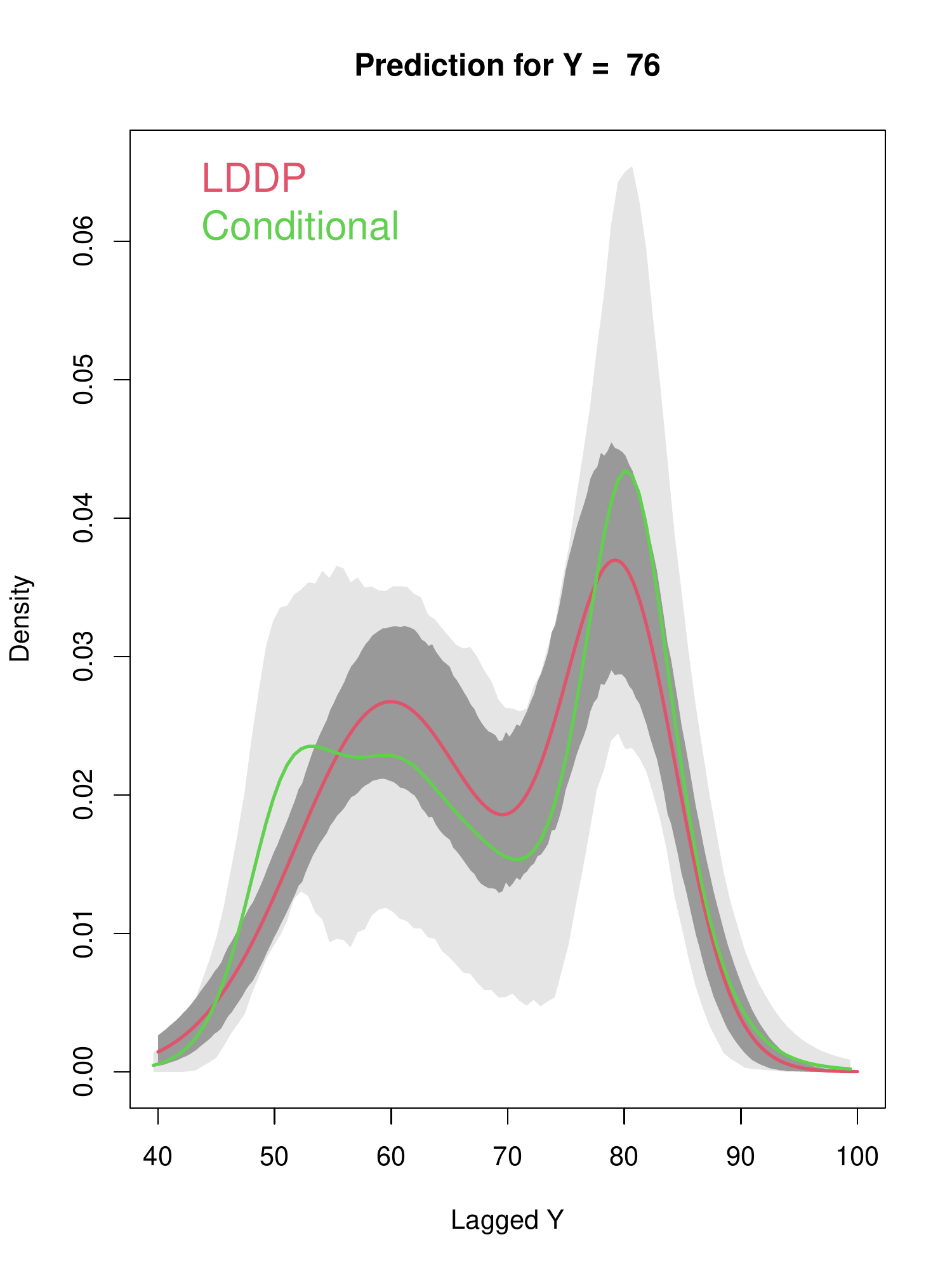}
}
\hspace{-0.5cm}\subfigure[] 
{
    \includegraphics[width=5.5cm]{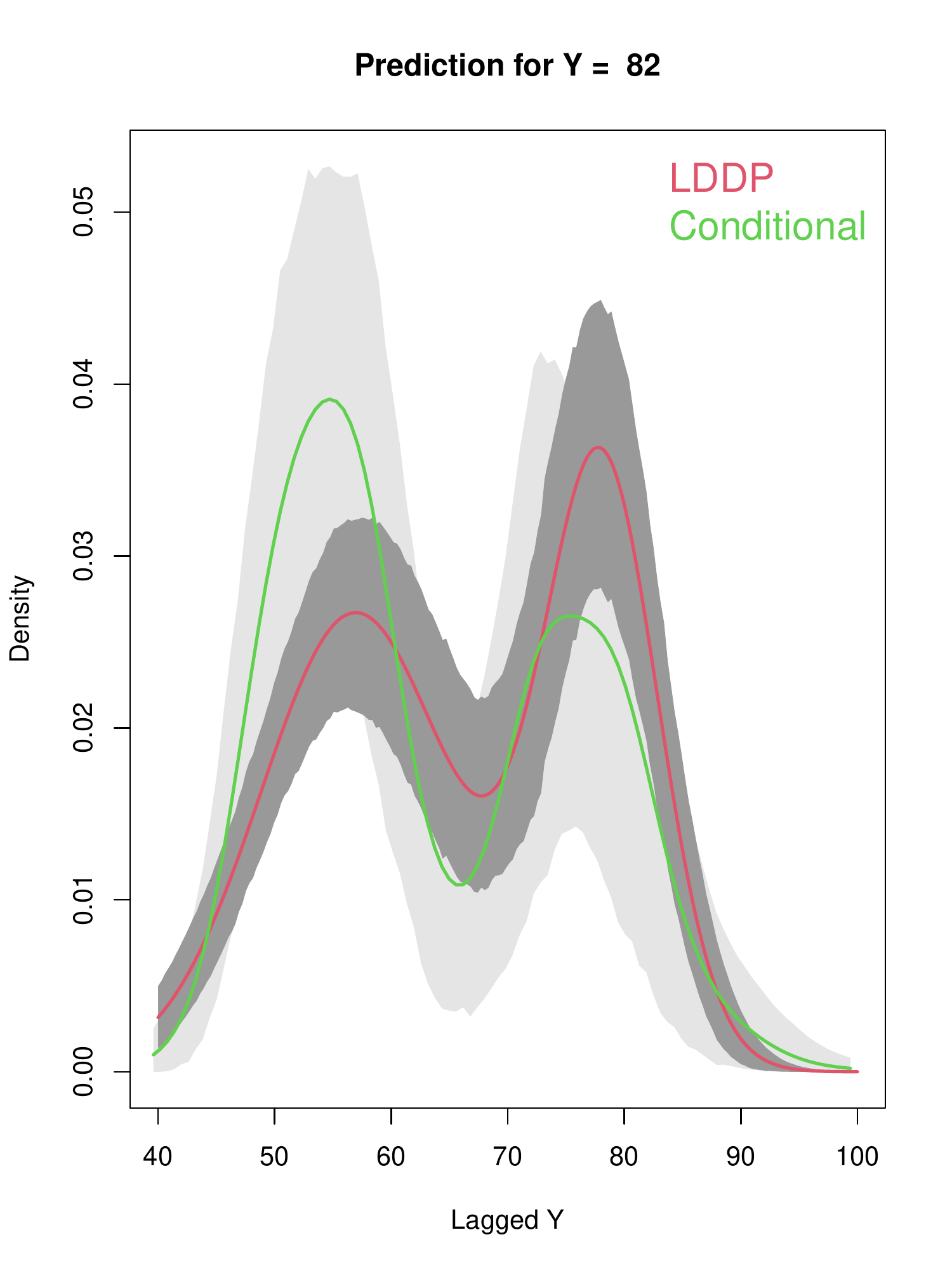}
}
\caption{Old Faithful Geyser data: 
  Posterior estimated $G_x$, i.e., 
  posterior predictive densities (mean and point-wise 
  95\% HPD intervals) for the waiting times at lagged times (a)
  $y_{t-1}=58$, (b) $y_{t-1}=76$ and (c) $y_{t-1}=82$, including 95\%
  HPD credibility bands. The red curve shows inference under the
  LDDP model. The green curve shows inference under the conditional
  density approach.}\label{fig:GEYSERcompare}
\end{figure}

\begin{figure}
\centering
 \includegraphics[height=6.5in,width=7in]{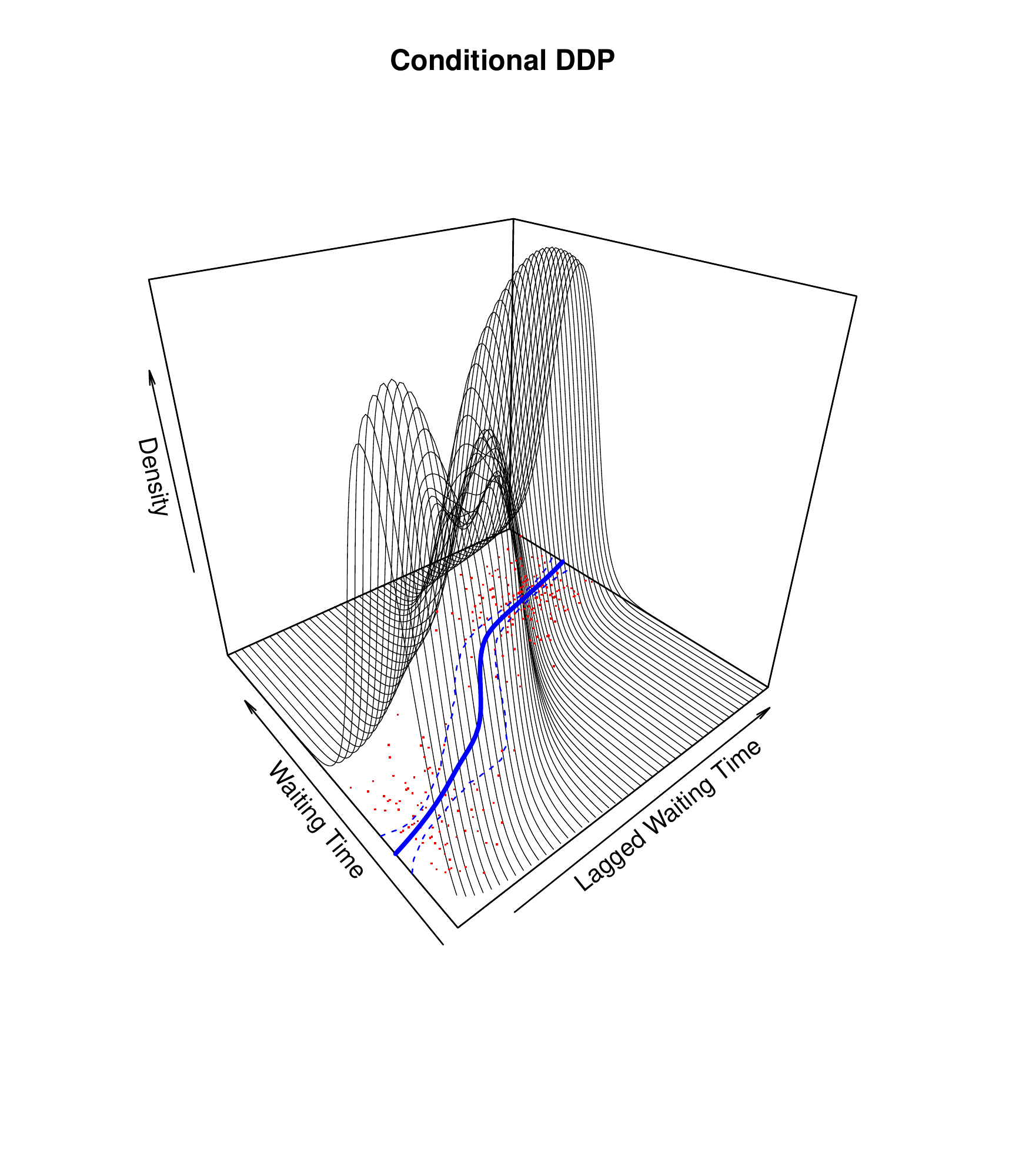}
 \caption{Old Faithful Geyser data:
    Posterior estimated densities $G_x$ for a grid of
   lagged values $x_t=y_{t-1}$. The blue curve shows
   conditional mean process (solid line), with 95\% credible intervals
   (dashed line).
 }\label{fig:perspGeyser}
\end{figure}

In all cases, we used hyperparameters as in
\cite{jara;hanson;quintana;mueller;rosner;2011}. Results for 
{\bf D1} are shown in Figure~\ref{fig:GEYSERcompare}.
In particular, we show a comparison posterior inference 
for $G_x$ for  (a) $y_{t-1}=58$, (b) $y_{t-1}=76$
and (c) $y_{t-1}=82$.
While there are some model-specific differences in the
 estimated distributions $G_x$,  
they both largely agree on the bimodal nature.

Figure~\ref{fig:perspGeyser} shows $G_x$ over a grid of lagged values
$x_t=y_{t-1}$, under the conditional density approach.
In this figure, the bimodality is also seen in the data (red dots).
The solid blue curve in Figure~\ref{fig:perspGeyser}
shows the posterior mean $E(y_t\mid y_{t-1})$ with 95\% credibility
bands (blue dashed curves).  

\begin{figure}
\centering
\subfigure[] 
{
\hspace{-2cm}    \includegraphics[width=4in]{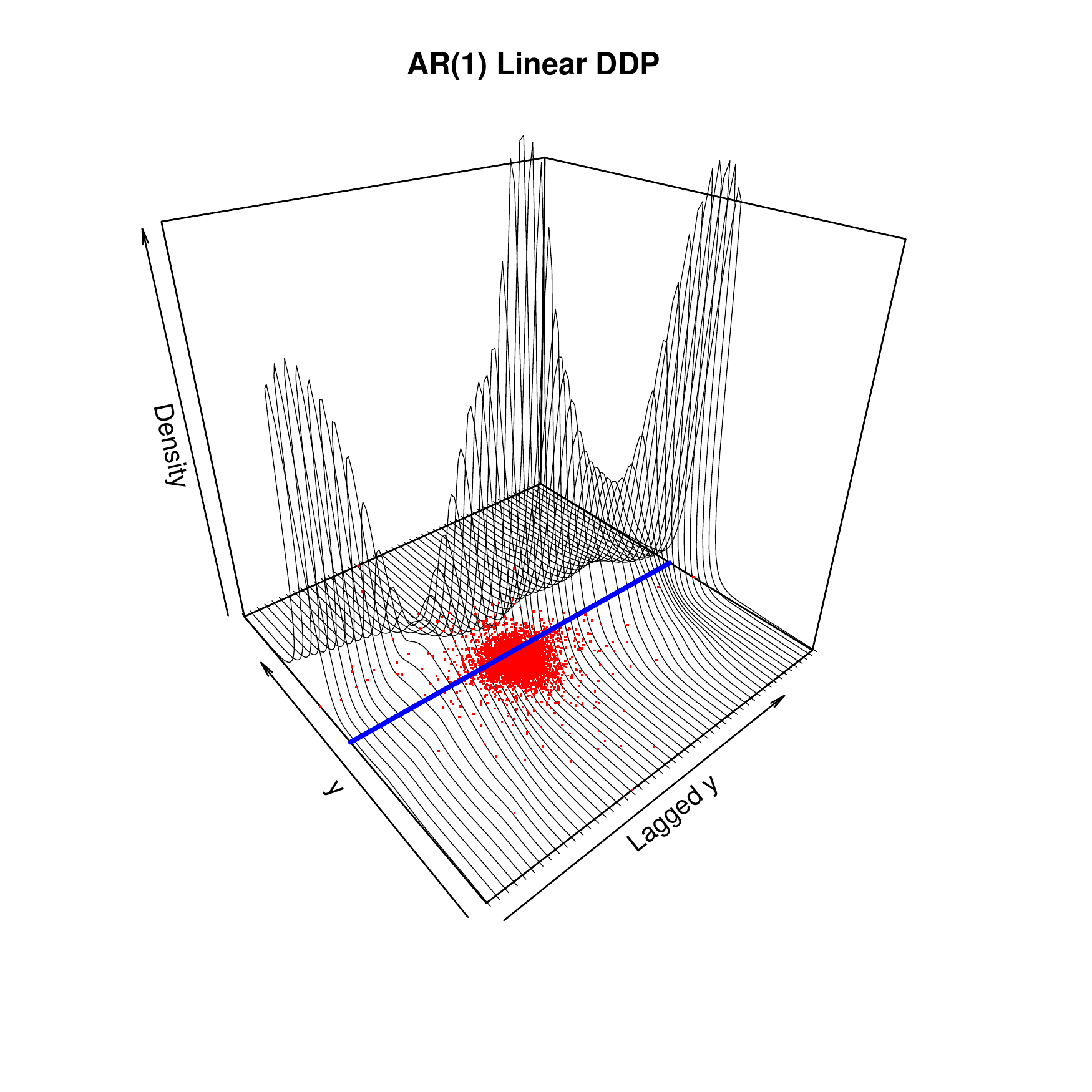}
}
\hspace{-2.5cm}\subfigure[] 
{
    \includegraphics[width=4in]{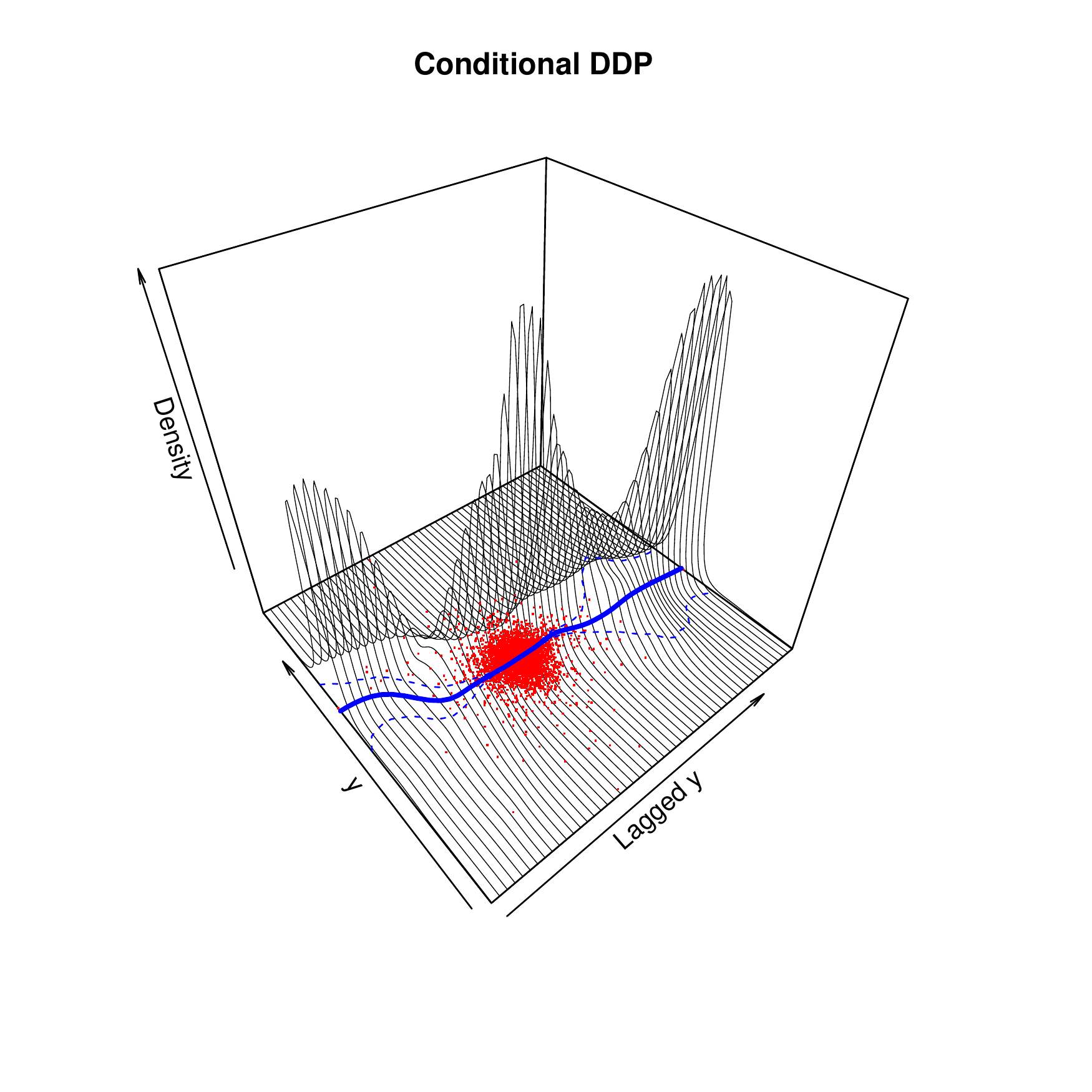}
}
\caption{S\&P500 data: Posterior  estimated densities $G_x$ for a
  grid of $x=y_{t-1}$. The blue lines show the conditional mean
  process, with dashed lines for 95\% HDP intervals.
  Panel (a) shows inference under the LDDP model, and panel (b) shows
  inference under the conditional density approach.}
\label{fig:SPCompare}
\end{figure}

In contrast, similar inference for the LDDP (not shown) shows a
straight line for the mean process $E(G_x | \yb)$, as a function of $x$,
as is implied by the linear structure of $\th_h(x)$ under the LDDP.  
See also Figure~\ref{fig:SPCompare} below

 Figure~\ref{fig:SPCompare} shows the same results for the S\&P500
data, using the same models as above.
As before, the data are shown as red dots, and blue lines show the
posterior predictive means (solid line) together with a 95\% HPD
interval (dashed line). 
Interestingly, for the LDDP model the HPD lines fall outside the
plotting region.  


\section{Concluding remarks}\label{sect:disc}


DDPs have come a long way since they were originally proposed. By its very
definition, a DDP has the potential to incorporate covariate indexing
(dependence) either in the atoms or the weights or both. The results in
\cite{barrientos;jara;quintana;2012} show that under full support of the
stochastic processes that are used to convey covariate dependence, the
resulting DDP has full support in the space $\FF=\{F_{\xb}:\,\xb\in\XX\}$.
This holds true for all of the basic DDP constructions: single-atoms,
single-weights, and with dependence in both. A natural question is then:
which DDP version is the best? There is no final answer to this question,
although DDPs with dependence in both atoms and weights are less commonly
found, mostly due to the computational complexity of implementation
entails. An exception to this is the conditional approach described
in Section~\ref{joint}. In broad terms, the single-weights models are
typically easier to fit, as the standard algorithms designed to implement
posterior simulation in the context of DPs can be applied with minor
adjustments. See, e.g., the computational aspects in
\cite{deiorio;mueller;rosner;maceachern;2004}. The same
applies for the LDDP.

On the other hand, the single-atoms
models are typically less  attractive  from a computational viewpoint,
mainly due to how covariate dependence is encoded in the definition of the
weight processes $\{w_h(\xb):\,\xb\in\XX\}$. However, the single-atoms DDP
allows  for the  prior probability distribution on the
partitions to  change with $\xb$, a feature that is not supported by the
single-weights DDP. For a formal description of this feature, let
$\GG=\{G_{\xb}:\, \xb\in\XX\}$ denote the family of random probability
measures with DDP prior, as before. Let $\rho_{\xb}$ denote the partition
of $\{1,\ldots,n\}$ that is implied by a hypothetical sample from
$G_{\xb}$, of size $n$. Under the single-weights DDP, $p(\rho_{\xb}\mid
\GG)$ is invariant across $\xb$; but not so under the single-atoms DDP.
This is the case since the prior on the random partition $\rho_{\xb}$ is
determined by the weights in $G_{\xb}$.

Models for dependent probability distributions do not easily allow for
the incorporation of existing prior information about arbitrary
functionals. A modeler is unlikely to have prior knowledge about all
aspects of a collection of probability measures, but could have real
historical prior information about specific functionals (such as the mean
or quantile functions). For example, such information could be obtained as
the product of applying parametric or (classical) nonparametric approaches
to previous data. Furthermore, even in models for single (non-dependent)
probability measures, 
 the derivation of  the induced
distribution for  arbitrary  functionals is challenging and,
thus,  usually not exploited.  
This makes the prior elicitation process 
difficult. We refer the reader to
\cite{lijoi;pruenster;2009} for an exhaustive summary of existing
results concerning distributional properties of functional of single and
discrete random probability measures.

In the context of a single probability measure,
\cite{kessler;hoff;dunson;2015} proposed a clever construction of a BNP
model with a given distribution on a finite set of functionals. Their
approach is based on the conditional distribution of a standard BNP prior,
given the functionals of interest. A Metropolis-Hastings MCMC algorithm is
proposed to explore the posterior distribution under the marginally
specified BNP model, where the standard BNP model is used as a candidate
generating model, and that is closely related to the well-known
importance-sampling approach for assessing prior sensitivity. Their MCMC
algorithm is developed for DP-based models and relies on the
marginalization of the random probability measure. Thus, a Monte Carlo
approximation of the functionals of interest is employed at any step of
the MCMC algorithm to obtain approximated posterior samples of the
functionals of interest. The study of extensions of the approach proposed
by \cite{kessler;hoff;dunson;2015} to the context of sets of
predictor-dependent probability measures is a topic of interest for future
research.

An interesting topic has been recently brought up
by~\cite{campbell2019local}. They introduced a relaxed version of the
notion of exchangeability, {\em local exchangeability}, which considers
bounded changes in total variation norm of the distribution of
observations under permutations of data having nearby covariate values.
This notion generalizes that of exchangeability and partial
exchangeability. The work by \cite{campbell2019local} discusses conditions
under which a version of de Finetti's Theorem holds in such a way that a
DDP is the corresponding de Finetti measure, i.e. conditional independence
of the observations under a DDP is still true. The study of extensions and
applications of these and related results is another topic of interest for
future research.

The bulk of work on the DDP and related methods focuses on the family of
conditional distributions $\GG=\{G_{\xb}:\,\xb\in\XX\}$ and models where
an observation $y$ is associated with a single value of the covariate
$\xb$.  When data are longitudinal, spatial or functional, the
observations may be considered to have dependence that cannot be captured
by the marginal distributions $G_{\xb}$. See, for
example,~\cite{xu;maceachern;xu;15} who separate dependence in financial
data series from the marginal distributions. Many open questions remain in
this direction.

Finally,  the idea of introducing dependence through normalization, e.g.
as mentioned earlier in Section~\ref{sect:other} can be further exploited
and extended to more general cases, including going beyond the context of
DDPs.


\section*{Acknowledgements}
A. Jara's and F. Quintana's research is supported by Millennium Science
Initiative of the Ministry of Economy, Development, and Tourism, grant
``Millennium Nucleus Center for the Discovery  of Structures in Complex
Data''. A. Jara is also supported by Fondecyt grant 1180640, F. Quintana
is also supported by Fondecyt grant 1180034. P. M\"uller acknowledges
partial support from grant
NSF/DMS 1952679 from the National Science Foundation, and under
R01 CA132897 from the U.S. National Cancer Institute.

\bibliographystyle{imsart-nameyear}

\begin{thebibliography}{85}

\bibitem[\protect\citeauthoryear{Aldous}{1985}]{aldous:85}
\begin{bincollection}[author]
\bauthor{\bsnm{Aldous},~\bfnm{David~J.}\binits{D.~J.}}
(\byear{1985}).
\btitle{Exchangeability and related topics}.
In \bbooktitle{\'{E}cole d'\'{e}t\'{e} de probabilit\'{e}s de {S}aint-{F}lour,
  {XIII}---1983}.
\bseries{Lecture Notes in Math.}
\bvolume{1117}
\bpages{1--198}.
\bpublisher{Springer, Berlin}.
\end{bincollection}
\endbibitem

\bibitem[\protect\citeauthoryear{Argiento, Guglielmi and
  Pievatolo}{2010}]{Argiento:2010}
\begin{barticle}[author]
\bauthor{\bsnm{Argiento},~\bfnm{Raffaele}\binits{R.}},
  \bauthor{\bsnm{Guglielmi},~\bfnm{Alessandra}\binits{A.}} \AND
  \bauthor{\bsnm{Pievatolo},~\bfnm{Antonio}\binits{A.}}
(\byear{2010}).
\btitle{{B}ayesian density estimation and model selection using nonparametric
  hierarchical mixtures}.
\bjournal{Computational Statistics \& Data Analysis}
\bvolume{54}
\bpages{816 - 832}.
\end{barticle}
\endbibitem

\bibitem[\protect\citeauthoryear{Ascolani, Lijoi and
  Ruggiero}{2020}]{ascolani2020}
\begin{barticle}[author]
\bauthor{\bsnm{Ascolani},~\bfnm{Filippo}\binits{F.}},
  \bauthor{\bsnm{Lijoi},~\bfnm{Antonio}\binits{A.}} \AND
  \bauthor{\bsnm{Ruggiero},~\bfnm{Matteo}\binits{M.}}
(\byear{2020}).
\btitle{Predictive inference with Fleming–Viot-driven dependent Dirichlet
  processes}.
\bjournal{Bayesian Analysis}.
\bnote{Advance publication}.
\bdoi{10.1214/20-BA1206}
\end{barticle}
\endbibitem

\bibitem[\protect\citeauthoryear{Barndorff-Nielsen}{1973}]{barndorff;1973}
\begin{barticle}[author]
\bauthor{\bsnm{Barndorff-Nielsen},~\bfnm{O.}\binits{O.}}
(\byear{1973}).
\btitle{On {$M$}-ancillarity}.
\bjournal{Biometrika}
\bvolume{60}
\bpages{447--455}.
\end{barticle}
\endbibitem

\bibitem[\protect\citeauthoryear{Barndorff-Nielsen}{1978}]{barndorff;1978}
\begin{bbook}[author]
\bauthor{\bsnm{Barndorff-Nielsen},~\bfnm{Ole}\binits{O.}}
(\byear{1978}).
\btitle{Information and exponential families in statistical theory}.
\bpublisher{John Wiley \& Sons, Ltd., Chichester}
\bnote{Wiley Series in Probability and Mathematical Statistics}.
\end{bbook}
\endbibitem

\bibitem[\protect\citeauthoryear{Barrientos, Jara and
  Quintana}{2012}]{barrientos;jara;quintana;2012}
\begin{barticle}[author]
\bauthor{\bsnm{Barrientos},~\bfnm{A~F}\binits{A.~F.}},
  \bauthor{\bsnm{Jara},~\bfnm{A}\binits{A.}} \AND
  \bauthor{\bsnm{Quintana},~\bfnm{F~A}\binits{F.~A.}}
(\byear{2012}).
\btitle{{On the support of MacEachern's dependent Drichlet processes and
  extensions}}.
\bjournal{Bayesian Analysis}
\bvolume{7}
\bpages{277-- 310}.
\end{barticle}
\endbibitem

\bibitem[\protect\citeauthoryear{Barrios et~al.}{2013}]{barriosNRMI2013}
\begin{barticle}[author]
\bauthor{\bsnm{Barrios},~\bfnm{Ernesto}\binits{E.}},
  \bauthor{\bsnm{Lijoi},~\bfnm{Antonio}\binits{A.}},
  \bauthor{\bsnm{Nieto-Barajas},~\bfnm{Luis~E.}\binits{L.~E.}} \AND
  \bauthor{\bsnm{Pr\"unster},~\bfnm{Igor}\binits{I.}}
(\byear{2013}).
\btitle{Modeling with normalized random measure mixture models}.
\bjournal{Statistical Science}
\bvolume{28}
\bpages{313--334}.
\bdoi{10.1214/13-STS416}
\end{barticle}
\endbibitem

\bibitem[\protect\citeauthoryear{Beraha, Guglielmi and
  Quintana}{2020}]{beraha2020semihierarchical}
\begin{bmisc}[author]
\bauthor{\bsnm{Beraha},~\bfnm{Mario}\binits{M.}},
  \bauthor{\bsnm{Guglielmi},~\bfnm{Alessandra}\binits{A.}} \AND
  \bauthor{\bsnm{Quintana},~\bfnm{Fernando~A.}\binits{F.~A.}}
(\byear{2020}).
\btitle{The semi-hierarchical Dirichlet Process and its application to
  clustering homogeneous distributions}.
\end{bmisc}
\endbibitem

\bibitem[\protect\citeauthoryear{Camerlenghi et~al.}{2019}]{camerlenghi2019}
\begin{barticle}[author]
\bauthor{\bsnm{Camerlenghi},~\bfnm{Federico}\binits{F.}},
  \bauthor{\bsnm{Dunson},~\bfnm{David~B.}\binits{D.~B.}},
  \bauthor{\bsnm{Lijoi},~\bfnm{Antonio}\binits{A.}},
  \bauthor{\bsnm{Pr\"unster},~\bfnm{Igor}\binits{I.}} \AND
  \bauthor{\bsnm{Rodr\'{\i}guez},~\bfnm{Abel}\binits{A.}}
(\byear{2019}).
\btitle{Latent Nested Nonparametric Priors (with Discussion)}.
\bjournal{Bayesian Anal.}
\bvolume{14}
\bpages{1303--1356}.
\bdoi{10.1214/19-BA1169}
\end{barticle}
\endbibitem

\bibitem[\protect\citeauthoryear{Campbell et~al.}{2019}]{campbell2019local}
\begin{bmisc}[author]
\bauthor{\bsnm{Campbell},~\bfnm{Trevor}\binits{T.}},
  \bauthor{\bsnm{Syed},~\bfnm{Saifuddin}\binits{S.}},
  \bauthor{\bsnm{Yang},~\bfnm{Chiao-Yu}\binits{C.-Y.}},
  \bauthor{\bsnm{Jordan},~\bfnm{Michael~I.}\binits{M.~I.}} \AND
  \bauthor{\bsnm{Broderick},~\bfnm{Tamara}\binits{T.}}
(\byear{2019}).
\btitle{Local Exchangeability}.
\end{bmisc}
\endbibitem

\bibitem[\protect\citeauthoryear{Caron et~al.}{2008}]{caronetal:08}
\begin{barticle}[author]
\bauthor{\bsnm{Caron},~\bfnm{F.}\binits{F.}},
  \bauthor{\bsnm{Davy},~\bfnm{M.}\binits{M.}},
  \bauthor{\bsnm{Doucet},~\bfnm{A.}\binits{A.}},
  \bauthor{\bsnm{Duflos},~\bfnm{E.}\binits{E.}} \AND
  \bauthor{\bsnm{Vanheeghe},~\bfnm{P.}\binits{P.}}
(\byear{2008}).
\btitle{{Bayesian Inference for Linear Dynamic Models with Dirichlet Process
  Mixtures}}.
\bjournal{IEEE Transactions on Signal Processing}
\bvolume{56}
\bpages{71--84}.
\end{barticle}
\endbibitem

\bibitem[\protect\citeauthoryear{Chen, Ding and
  Buntine}{2012}]{ICML2012Chen_476}
\begin{binproceedings}[author]
\bauthor{\bsnm{Chen},~\bfnm{Changyou}\binits{C.}},
  \bauthor{\bsnm{Ding},~\bfnm{Nan}\binits{N.}} \AND
  \bauthor{\bsnm{Buntine},~\bfnm{Wray}\binits{W.}}
(\byear{2012}).
\btitle{Dependent Hierarchical Normalized Random Measures for Dynamic Topic
  Modeling}.
In \bbooktitle{Proceedings of the 29th International Conference on Machine
  Learning (ICML-12)}
(\beditor{\bfnm{John}\binits{J.}~\bsnm{Langford}} \AND
  \beditor{\bfnm{Joelle}\binits{J.}~\bsnm{Pineau}}, eds.).
\bseries{ICML '12}
\bpages{895--902}.
\bpublisher{Omnipress}, \baddress{New York, NY, USA}.
\end{binproceedings}
\endbibitem

\bibitem[\protect\citeauthoryear{Chen et~al.}{2013}]{pmlr-v28-chen13i}
\begin{binproceedings}[author]
\bauthor{\bsnm{Chen},~\bfnm{Changyou}\binits{C.}},
  \bauthor{\bsnm{Rao},~\bfnm{Vinayak}\binits{V.}},
  \bauthor{\bsnm{Buntine},~\bfnm{Wray}\binits{W.}} \AND
  \bauthor{\bsnm{Teh},~\bfnm{Yee~Whye}\binits{Y.~W.}}
(\byear{2013}).
\btitle{Dependent Normalized Random Measures}.
In \bbooktitle{Proceedings of the 30th International Conference on Machine
  Learning}
(\beditor{\bfnm{Sanjoy}\binits{S.}~\bsnm{Dasgupta}} \AND
  \beditor{\bfnm{David}\binits{D.}~\bsnm{McAllester}}, eds.).
\bseries{Proceedings of Machine Learning Research}
\bvolume{28}
\bpages{969--977}.
\bpublisher{PMLR}, \baddress{Atlanta, Georgia, USA}.
\end{binproceedings}
\endbibitem

\bibitem[\protect\citeauthoryear{Chipman, George and
  McCulloch}{2010}]{chipman2010}
\begin{barticle}[author]
\bauthor{\bsnm{Chipman},~\bfnm{Hugh~A.}\binits{H.~A.}},
  \bauthor{\bsnm{George},~\bfnm{Edward~I.}\binits{E.~I.}} \AND
  \bauthor{\bsnm{McCulloch},~\bfnm{Robert~E.}\binits{R.~E.}}
(\byear{2010}).
\btitle{BART: Bayesian additive regression trees}.
\bjournal{Ann. Appl. Stat.}
\bvolume{4}
\bpages{266--298}.
\bdoi{10.1214/09-AOAS285}
\end{barticle}
\endbibitem

\bibitem[\protect\citeauthoryear{Chung and Dunson}{2009}]{chung;dunson;2009}
\begin{barticle}[author]
\bauthor{\bsnm{Chung},~\bfnm{Y}\binits{Y.}} \AND
  \bauthor{\bsnm{Dunson},~\bfnm{D~B}\binits{D.~B.}}
(\byear{2009}).
\btitle{{Nonparametric Bayes conditional distribution modeling with variable
  selection}}.
\bjournal{Journal of the American Statistical Association}
\bvolume{104}
\bpages{1646--1660}.
\end{barticle}
\endbibitem

\bibitem[\protect\citeauthoryear{Cifarelli and
  Regazzini}{1978}]{cifarelli;regazzini;78}
\begin{btechreport}[author]
\bauthor{\bsnm{Cifarelli},~\bfnm{D}\binits{D.}} \AND
  \bauthor{\bsnm{Regazzini},~\bfnm{E}\binits{E.}}
(\byear{1978}).
\btitle{{Problemi statistici non parametrici in condizioni di scambialbilita
  parziale e impiego di medie associative}}
\btype{Technical Report},
\bpublisher{Quaderni Istituto Matematica Finanziaria, Torino}.
\end{btechreport}
\endbibitem

\bibitem[\protect\citeauthoryear{De~Iorio
  et~al.}{2004}]{deiorio;mueller;rosner;maceachern;2004}
\begin{barticle}[author]
\bauthor{\bsnm{De~Iorio},~\bfnm{M}\binits{M.}},
  \bauthor{\bsnm{M\"{u}ller},~\bfnm{P}\binits{P.}},
  \bauthor{\bsnm{Rosner},~\bfnm{G~L}\binits{G.~L.}} \AND
  \bauthor{\bsnm{MacEachern},~\bfnm{S~N}\binits{S.~N.}}
(\byear{2004}).
\btitle{{An ANOVA model for dependent random measures}}.
\bjournal{Journal of the American Statistical Association}
\bvolume{99}
\bpages{205--215}.
\end{barticle}
\endbibitem

\bibitem[\protect\citeauthoryear{De~Iorio
  et~al.}{2009}]{deiorio;johnson;mueller;rosner;2009}
\begin{barticle}[author]
\bauthor{\bsnm{De~Iorio},~\bfnm{M}\binits{M.}},
  \bauthor{\bsnm{Johnson},~\bfnm{W~O}\binits{W.~O.}},
  \bauthor{\bsnm{M\"{u}ller},~\bfnm{P}\binits{P.}} \AND
  \bauthor{\bsnm{Rosner},~\bfnm{G~L}\binits{G.~L.}}
(\byear{2009}).
\btitle{{Bayesian nonparametric non-proportional hazards survival modelling}}.
\bjournal{Biometrics}
\bvolume{65}
\bpages{762--771}.
\end{barticle}
\endbibitem

\bibitem[\protect\citeauthoryear{De~la Cruz-Mes{\'{\i}}a, Quintana and
  M{\"u}ller}{2007}]{DEQUMU:07}
\begin{barticle}[author]
\bauthor{\bparticle{De~la}
  \bsnm{Cruz-Mes{\'{\i}}a},~\bfnm{Rolando}\binits{R.}},
  \bauthor{\bsnm{Quintana},~\bfnm{Fernando~A.}\binits{F.~A.}} \AND
  \bauthor{\bsnm{M{\"u}ller},~\bfnm{Peter}\binits{P.}}
(\byear{2007}).
\btitle{Semiparametric {B}ayesian classification with longitudinal markers}.
\bjournal{Journal of the Royal Statistical Society. Series C. Applied
  Statistics}
\bvolume{56}
\bpages{119--137}.
\end{barticle}
\endbibitem

\bibitem[\protect\citeauthoryear{Devroye}{1986}]{devroye:1986}
\begin{bbook}[author]
\bauthor{\bsnm{Devroye},~\bfnm{Luc}\binits{L.}}
(\byear{1986}).
\btitle{Non-Uniform Random Variate Generation(originally published with}.
\bpublisher{Springer-Verlag}.
\end{bbook}
\endbibitem

\bibitem[\protect\citeauthoryear{Di~Lucca et~al.}{2012}]{diluccaetal:12}
\begin{barticle}[author]
\bauthor{\bsnm{Di~Lucca},~\bfnm{M.~A.}\binits{M.~A.}},
  \bauthor{\bsnm{Guglielmi},~\bfnm{A.}\binits{A.}},
  \bauthor{\bsnm{M\"uller},~\bfnm{P.}\binits{P.}} \AND
  \bauthor{\bsnm{Quintana},~\bfnm{F.~A.}\binits{F.~A.}}
(\byear{2012}).
\btitle{{A simple class of Bayesian nonparametric autoregression models}}.
\bjournal{Bayesian Analysis}
\bvolume{8}
\bpages{63--88}.
\end{barticle}
\endbibitem

\bibitem[\protect\citeauthoryear{Duan, Guindani and
  Gelfand}{2007}]{duan;guidani;gelfand;2007}
\begin{barticle}[author]
\bauthor{\bsnm{Duan},~\bfnm{J~A}\binits{J.~A.}},
  \bauthor{\bsnm{Guindani},~\bfnm{M}\binits{M.}} \AND
  \bauthor{\bsnm{Gelfand},~\bfnm{A~E}\binits{A.~E.}}
(\byear{2007}).
\btitle{{Generalized spatial Dirichlet process models}}.
\bjournal{Biometrika}
\bvolume{94}
\bpages{809--825}.
\end{barticle}
\endbibitem

\bibitem[\protect\citeauthoryear{Dunson and
  Herring}{2006}]{dunson;herring;2006}
\begin{btechreport}[author]
\bauthor{\bsnm{Dunson},~\bfnm{D~B}\binits{D.~B.}} \AND
  \bauthor{\bsnm{Herring},~\bfnm{A~H}\binits{A.~H.}}
(\byear{2006}).
\btitle{{Semiparametric Bayesian latent trajectory models}}
\btype{Technical Report},
\bpublisher{ISDS Discussion Paper 16, Duke University, Durham, NC, USA.}
\end{btechreport}
\endbibitem

\bibitem[\protect\citeauthoryear{Dunson and Park}{2008}]{dunson;park;2008}
\begin{barticle}[author]
\bauthor{\bsnm{Dunson},~\bfnm{D~B}\binits{D.~B.}} \AND
  \bauthor{\bsnm{Park},~\bfnm{J~H}\binits{J.~H.}}
(\byear{2008}).
\btitle{{Kernel stick-breaking processes}}.
\bjournal{Biometrika}
\bvolume{95}
\bpages{307--323}.
\end{barticle}
\endbibitem

\bibitem[\protect\citeauthoryear{Dunson, Pillai and
  Park}{2007}]{dunson;pillai;park;2007}
\begin{barticle}[author]
\bauthor{\bsnm{Dunson},~\bfnm{D~B}\binits{D.~B.}},
  \bauthor{\bsnm{Pillai},~\bfnm{N}\binits{N.}} \AND
  \bauthor{\bsnm{Park},~\bfnm{J~H}\binits{J.~H.}}
(\byear{2007}).
\btitle{{Bayesian density regression}}.
\bjournal{Journal of the Royal Statistical Society, Series B}
\bvolume{69}
\bpages{163--183}.
\end{barticle}
\endbibitem

\bibitem[\protect\citeauthoryear{Epifani and Lijoi}{2010}]{epifani;lijoi;2010}
\begin{barticle}[author]
\bauthor{\bsnm{Epifani},~\bfnm{I}\binits{I.}} \AND
  \bauthor{\bsnm{Lijoi},~\bfnm{A}\binits{A.}}
(\byear{2010}).
\btitle{{Nonparametric priors for vectors of survival functions}}.
\bjournal{Statistica Sinica}
\bvolume{20}
\bpages{1455--1484}.
\end{barticle}
\endbibitem

\bibitem[\protect\citeauthoryear{Faraway}{2016}]{faraway:16}
\begin{bbook}[author]
\bauthor{\bsnm{Faraway},~\bfnm{Julian~J.}\binits{J.~J.}}
(\byear{2016}).
\btitle{Extending the linear model with {R}}.
\bseries{Chapman \& Hall/CRC Texts in Statistical Science Series}.
\bpublisher{CRC Press, Boca Raton, FL}
\bnote{Generalized linear, mixed effects and nonparametric regression models,
  Second edition [of MR2192856]}.
\end{bbook}
\endbibitem

\bibitem[\protect\citeauthoryear{Favaro and Teh}{2013}]{favaro2013}
\begin{barticle}[author]
\bauthor{\bsnm{Favaro},~\bfnm{Stefano}\binits{S.}} \AND
  \bauthor{\bsnm{Teh},~\bfnm{Yee~Whye}\binits{Y.~W.}}
(\byear{2013}).
\btitle{MCMC for Normalized Random Measure Mixture Models}.
\bjournal{Statistical Science}
\bvolume{28}
\bpages{335--359}.
\bdoi{10.1214/13-STS422}
\end{barticle}
\endbibitem

\bibitem[\protect\citeauthoryear{Ferguson}{1973}]{ferguson;73}
\begin{barticle}[author]
\bauthor{\bsnm{Ferguson},~\bfnm{T~S}\binits{T.~S.}}
(\byear{1973}).
\btitle{A {B}ayesian analysis of some nonparametric problems}.
\bjournal{Annals of Statistics}
\bvolume{1}
\bpages{209--230}.
\end{barticle}
\endbibitem

\bibitem[\protect\citeauthoryear{Ferguson}{1974}]{ferguson;74}
\begin{barticle}[author]
\bauthor{\bsnm{Ferguson},~\bfnm{T~S}\binits{T.~S.}}
(\byear{1974}).
\btitle{Prior distribution on the spaces of probability measures}.
\bjournal{Annals of Statistics}
\bvolume{2}
\bpages{615--629}.
\end{barticle}
\endbibitem

\bibitem[\protect\citeauthoryear{Fuentes-Garc\'ia, Mena and
  Walker}{2009}]{fuentes;mena;walker;2009}
\begin{barticle}[author]
\bauthor{\bsnm{Fuentes-Garc\'ia},~\bfnm{R}\binits{R.}},
  \bauthor{\bsnm{Mena},~\bfnm{R}\binits{R.}} \AND
  \bauthor{\bsnm{Walker},~\bfnm{S~G}\binits{S.~G.}}
(\byear{2009}).
\btitle{{A nonparametric dependent process for Bayesian regression}}.
\bjournal{Statistics and Probability Letters}
\bvolume{79}
\bpages{1112--1119}.
\end{barticle}
\endbibitem

\bibitem[\protect\citeauthoryear{Gelfand and
  Kottas}{2001}]{gelfand;kottas;2001}
\begin{barticle}[author]
\bauthor{\bsnm{Gelfand},~\bfnm{A~E}\binits{A.~E.}} \AND
  \bauthor{\bsnm{Kottas},~\bfnm{A}\binits{A.}}
(\byear{2001}).
\btitle{{Nonparametric Bayesian modeling for stochastic order}}.
\bjournal{Annals of the Institute of Statistical Mathematics}
\bvolume{53}
\bpages{865--876}.
\end{barticle}
\endbibitem

\bibitem[\protect\citeauthoryear{Gelfand, Kottas and
  MacEachern}{2005}]{gelfand;kottas;maceachern;2005}
\begin{barticle}[author]
\bauthor{\bsnm{Gelfand},~\bfnm{A~E}\binits{A.~E.}},
  \bauthor{\bsnm{Kottas},~\bfnm{A}\binits{A.}} \AND
  \bauthor{\bsnm{MacEachern},~\bfnm{S~N}\binits{S.~N.}}
(\byear{2005}).
\btitle{{Bayesian nonparametric spatial modeling with Dirichlet process
  mixing}}.
\bjournal{Journal of the American Statistical Association}
\bvolume{100}
\bpages{1021--1035}.
\end{barticle}
\endbibitem

\bibitem[\protect\citeauthoryear{Giudici, Mezzetti and
  Muliere}{2003}]{giudici;mezzetti;muliere;2003}
\begin{barticle}[author]
\bauthor{\bsnm{Giudici},~\bfnm{P}\binits{P.}},
  \bauthor{\bsnm{Mezzetti},~\bfnm{M}\binits{M.}} \AND
  \bauthor{\bsnm{Muliere},~\bfnm{P}\binits{P.}}
(\byear{2003}).
\btitle{{Mixtures of Dirichlet process priors for variable selection in
  survival analysis}}.
\bjournal{Journal of Statistical Planning and Inference}
\bvolume{111}
\bpages{101--115}.
\end{barticle}
\endbibitem

\bibitem[\protect\citeauthoryear{Green and
  Richardson}{2001}]{green-richardson:2001}
\begin{barticle}[author]
\bauthor{\bsnm{Green},~\bfnm{Peter~J.}\binits{P.~J.}} \AND
  \bauthor{\bsnm{Richardson},~\bfnm{Sylvia}\binits{S.}}
(\byear{2001}).
\btitle{Modelling heterogeneity with and without the {D}irichlet process}.
\bjournal{Scandinavian Journal of Statistics. Theory and Applications}
\bvolume{28}
\bpages{355--375}.
\end{barticle}
\endbibitem

\bibitem[\protect\citeauthoryear{Griffin and Steel}{2006}]{griffin;steel;2006}
\begin{barticle}[author]
\bauthor{\bsnm{Griffin},~\bfnm{J~E}\binits{J.~E.}} \AND
  \bauthor{\bsnm{Steel},~\bfnm{M~F~J}\binits{M.~F.~J.}}
(\byear{2006}).
\btitle{{Order-based dependent Dirichlet processes}}.
\bjournal{Journal of the American Statistical Association}
\bvolume{101}
\bpages{179--194}.
\end{barticle}
\endbibitem

\bibitem[\protect\citeauthoryear{Griffin and Steel}{2010}]{griffin&steel:10}
\begin{barticle}[author]
\bauthor{\bsnm{Griffin},~\bfnm{J.~E.}\binits{J.~E.}} \AND
  \bauthor{\bsnm{Steel},~\bfnm{M.~F.~J.}\binits{M.~F.~J.}}
(\byear{2010}).
\btitle{Bayesian nonparametric modelling with the {D}irichlet process
  regression smoother}.
\bjournal{Statistica Sinica}
\bvolume{20}
\bpages{1507--1527}.
\end{barticle}
\endbibitem

\bibitem[\protect\citeauthoryear{Griffin and Steel}{2011}]{griffin&steel:11}
\begin{barticle}[author]
\bauthor{\bsnm{Griffin},~\bfnm{J.~E.}\binits{J.~E.}} \AND
  \bauthor{\bsnm{Steel},~\bfnm{M.~F.~J.}\binits{M.~F.~J.}}
(\byear{2011}).
\btitle{Stick-breaking autoregressive processes}.
\bjournal{Journal of Econometrics}
\bvolume{162}
\bpages{383--396}.
\end{barticle}
\endbibitem

\bibitem[\protect\citeauthoryear{Guti{\'e}rrez, Mena and
  Ruggiero}{2016}]{gutierrez2016time}
\begin{barticle}[author]
\bauthor{\bsnm{Guti{\'e}rrez},~\bfnm{Luis}\binits{L.}},
  \bauthor{\bsnm{Mena},~\bfnm{Rams{\'e}s~H}\binits{R.~H.}} \AND
  \bauthor{\bsnm{Ruggiero},~\bfnm{Matteo}\binits{M.}}
(\byear{2016}).
\btitle{{A time dependent Bayesian nonparametric model for air quality
  analysis}}.
\bjournal{Computational Statistics \& Data Analysis}
\bvolume{95}
\bpages{161--175}.
\end{barticle}
\endbibitem

\bibitem[\protect\citeauthoryear{Guti\'errez et~al.}{2019}]{gutierrez2019}
\begin{barticle}[author]
\bauthor{\bsnm{Guti\'errez},~\bfnm{Luis}\binits{L.}},
  \bauthor{\bsnm{Barrientos},~\bfnm{Andr\'es~F.}\binits{A.~F.}},
  \bauthor{\bsnm{Gonz\'alez},~\bfnm{Jorge}\binits{J.}} \AND
  \bauthor{\bsnm{Taylor-Rodr\'{\i}guez},~\bfnm{Daniel}\binits{D.}}
(\byear{2019}).
\btitle{A Bayesian Nonparametric Multiple Testing Procedure for Comparing
  Several Treatments Against a Control}.
\bjournal{Bayesian Analysis}
\bvolume{14}
\bpages{649--675}.
\bdoi{10.1214/18-BA1122}
\end{barticle}
\endbibitem

\bibitem[\protect\citeauthoryear{Gy\"{o}rfi et~al.}{2002}]{gyorfietal:02}
\begin{bbook}[author]
\bauthor{\bsnm{Gy\"{o}rfi},~\bfnm{L\'{a}szl\'{o}}\binits{L.}},
  \bauthor{\bsnm{Kohler},~\bfnm{Michael}\binits{M.}},
  \bauthor{\bsnm{Krzy\.~{z}ak},~\bfnm{Adam}\binits{A.}} \AND
  \bauthor{\bsnm{Walk},~\bfnm{Harro}\binits{H.}}
(\byear{2002}).
\btitle{A distribution-free theory of nonparametric regression}.
\bseries{Springer Series in Statistics}.
\bpublisher{Springer-Verlag, New York}.
\bdoi{10.1007/b97848}
\end{bbook}
\endbibitem

\bibitem[\protect\citeauthoryear{H\"{a}rdle}{1991}]{hardle:91}
\begin{bbook}[author]
\bauthor{\bsnm{H\"{a}rdle},~\bfnm{Wolfgang}\binits{W.}}
(\byear{1991}).
\btitle{Smoothing techniques}.
\bseries{Springer Series in Statistics}.
\bpublisher{Springer-Verlag, New York}
\bnote{With implementation in S}.
\end{bbook}
\endbibitem

\bibitem[\protect\citeauthoryear{James, Lijoi and
  Pr\"unster}{2009}]{james-lijoi-pruenster:09}
\begin{barticle}[author]
\bauthor{\bsnm{James},~\bfnm{Lancelot~F.}\binits{L.~F.}},
  \bauthor{\bsnm{Lijoi},~\bfnm{Antonio}\binits{A.}} \AND
  \bauthor{\bsnm{Pr\"unster},~\bfnm{Igor}\binits{I.}}
(\byear{2009}).
\btitle{Posterior Analysis for Normalized Random Measures with Independent
  Increments}.
\bjournal{Scandinavian Journal of Statistics}
\bvolume{36}
\bpages{76-97}.
\bdoi{10.1111/j.1467-9469.2008.00609.x}
\end{barticle}
\endbibitem

\bibitem[\protect\citeauthoryear{Jara and Hanson}{2011}]{jara;hanson;2011}
\begin{barticle}[author]
\bauthor{\bsnm{Jara},~\bfnm{A}\binits{A.}} \AND
  \bauthor{\bsnm{Hanson},~\bfnm{T}\binits{T.}}
(\byear{2011}).
\btitle{{A class of mixtures of dependent tail-free processes}}.
\bjournal{Biometrika}
\bvolume{98}
\bpages{553--566}.
\end{barticle}
\endbibitem

\bibitem[\protect\citeauthoryear{Jara
  et~al.}{2010}]{jara;lesaffre;deiorio;quintana;2010}
\begin{barticle}[author]
\bauthor{\bsnm{Jara},~\bfnm{A}\binits{A.}},
  \bauthor{\bsnm{Lesaffre},~\bfnm{E}\binits{E.}},
  \bauthor{\bsnm{De~Iorio},~\bfnm{M}\binits{M.}} \AND
  \bauthor{\bsnm{Quintana},~\bfnm{F~A}\binits{F.~A.}}
(\byear{2010}).
\btitle{{Bayesian semiparametric inference for multivariate
  doubly-interval-censored data}}.
\bjournal{The Annals of Applied Statistics}
\bvolume{4}
\bpages{2126--2149}.
\end{barticle}
\endbibitem

\bibitem[\protect\citeauthoryear{Jara
  et~al.}{2011}]{jara;hanson;quintana;mueller;rosner;2011}
\begin{barticle}[author]
\bauthor{\bsnm{Jara},~\bfnm{A}\binits{A.}},
  \bauthor{\bsnm{Hanson},~\bfnm{T}\binits{T.}},
  \bauthor{\bsnm{Quintana},~\bfnm{F}\binits{F.}},
  \bauthor{\bsnm{M\"{u}ller},~\bfnm{P}\binits{P.}} \AND
  \bauthor{\bsnm{Rosner},~\bfnm{G~L}\binits{G.~L.}}
(\byear{2011}).
\btitle{{DPpackage: Bayesian Semi- and Nonparametric Modeling in R}}.
\bjournal{Journal of Statistical Software}
\bvolume{40}
\bpages{1--30}.
\end{barticle}
\endbibitem

\bibitem[\protect\citeauthoryear{Kessler, Hoff and
  Dunson}{2015}]{kessler;hoff;dunson;2015}
\begin{barticle}[author]
\bauthor{\bsnm{Kessler},~\bfnm{D}\binits{D.}},
  \bauthor{\bsnm{Hoff},~\bfnm{P}\binits{P.}} \AND
  \bauthor{\bsnm{Dunson},~\bfnm{D}\binits{D.}}
(\byear{2015}).
\btitle{{Marginally specified priors for non-parametric Bayesian estimation}}.
\bjournal{Journal of the Royal Statistical Society, Series B}
\bvolume{77}
\bpages{35--58}.
\end{barticle}
\endbibitem

\bibitem[\protect\citeauthoryear{Klemel\"{a}}{2014}]{klemela:14}
\begin{bbook}[author]
\bauthor{\bsnm{Klemel\"{a}},~\bfnm{Jussi}\binits{J.}}
(\byear{2014}).
\btitle{Multivariate nonparametric regression and visualization}.
\bseries{Wiley Series in Computational Statistics}.
\bpublisher{John Wiley \& Sons, Inc., Hoboken, NJ}
\bnote{With R and applications to finance}.
\end{bbook}
\endbibitem

\bibitem[\protect\citeauthoryear{Kolossiatis, Griffin and
  Steel}{2013}]{kolossiatis&al:13}
\begin{barticle}[author]
\bauthor{\bsnm{Kolossiatis},~\bfnm{M.}\binits{M.}},
  \bauthor{\bsnm{Griffin},~\bfnm{J.~E.}\binits{J.~E.}} \AND
  \bauthor{\bsnm{Steel},~\bfnm{M.~F.~J.}\binits{M.~F.~J.}}
(\byear{2013}).
\btitle{On {B}ayesian nonparametric modelling of two correlated distributions}.
\bjournal{Statistics and Computing}
\bvolume{23}
\bpages{1--15}.
\bdoi{10.1007/s11222-011-9283-7}
\bmrnumber{3018346}
\end{barticle}
\endbibitem

\bibitem[\protect\citeauthoryear{Lau and So}{2008}]{lau&so:08}
\begin{barticle}[author]
\bauthor{\bsnm{Lau},~\bfnm{John~W.}\binits{J.~W.}} \AND
  \bauthor{\bsnm{So},~\bfnm{Mike K.~P.}\binits{M.~K.~P.}}
(\byear{2008}).
\btitle{{Bayesian mixture of autoregressive models}}.
\bjournal{Computational Statistics and Data Analysis}
\bvolume{53}
\bpages{38--60}.
\end{barticle}
\endbibitem

\bibitem[\protect\citeauthoryear{Lavine}{1992}]{lavine;92}
\begin{barticle}[author]
\bauthor{\bsnm{Lavine},~\bfnm{M}\binits{M.}}
(\byear{1992}).
\btitle{Some aspects of {P}olya tree distributions for statistical modeling}.
\bjournal{The Annals of Statistics}
\bvolume{20}
\bpages{1222--1235}.
\end{barticle}
\endbibitem

\bibitem[\protect\citeauthoryear{Leisen and Lijoi}{2011}]{leisen;lijoi;2011}
\begin{barticle}[author]
\bauthor{\bsnm{Leisen},~\bfnm{F}\binits{F.}} \AND
  \bauthor{\bsnm{Lijoi},~\bfnm{A}\binits{A.}}
(\byear{2011}).
\btitle{{Vectors of two--parameter Poisson--Dirichlet processes}}.
\bjournal{Journal of Multivariate Analysis}
\bvolume{102}
\bpages{482--495}.
\end{barticle}
\endbibitem

\bibitem[\protect\citeauthoryear{Lijoi, Nipoti and
  Pr\"{u}nster}{2014}]{lijoi;nipoti;pruenster;2014}
\begin{barticle}[author]
\bauthor{\bsnm{Lijoi},~\bfnm{A}\binits{A.}},
  \bauthor{\bsnm{Nipoti},~\bfnm{B}\binits{B.}} \AND
  \bauthor{\bsnm{Pr\"{u}nster},~\bfnm{I}\binits{I.}}
(\byear{2014}).
\btitle{{Bayesian inference with dependent normalized completely random
  measures}}.
\bjournal{Bernoulli}
\bvolume{20}
\bpages{1260--1291}.
\end{barticle}
\endbibitem

\bibitem[\protect\citeauthoryear{Lijoi and
  Pr\"{u}nster}{2009}]{lijoi;pruenster;2009}
\begin{barticle}[author]
\bauthor{\bsnm{Lijoi},~\bfnm{A}\binits{A.}} \AND
  \bauthor{\bsnm{Pr\"{u}nster},~\bfnm{I}\binits{I.}}
(\byear{2009}).
\btitle{{Distributional properties of means of random probability measures}}.
\bjournal{Statistical Surveys}
\bvolume{3}
\bpages{47–95}.
\end{barticle}
\endbibitem

\bibitem[\protect\citeauthoryear{Lin, Grimson and Fisher}{2010}]{NIPS2010_4151}
\begin{bincollection}[author]
\bauthor{\bsnm{Lin},~\bfnm{Dahua}\binits{D.}},
  \bauthor{\bsnm{Grimson},~\bfnm{Eric}\binits{E.}} \AND
  \bauthor{\bsnm{Fisher},~\bfnm{John~W.}\binits{J.~W.}}
(\byear{2010}).
\btitle{Construction of Dependent Dirichlet Processes based on Poisson
  Processes}.
In \bbooktitle{Advances in Neural Information Processing Systems 23}
(\beditor{\bfnm{J.~D.}\binits{J.~D.}~\bsnm{Lafferty}},
  \beditor{\bfnm{C.~K.~I.}\binits{C.~K.~I.}~\bsnm{Williams}},
  \beditor{\bfnm{J.}\binits{J.}~\bsnm{Shawe-Taylor}},
  \beditor{\bfnm{R.~S.}\binits{R.~S.}~\bsnm{Zemel}} \AND
  \beditor{\bfnm{A.}\binits{A.}~\bsnm{Culotta}}, eds.)
\bpages{1396--1404}.
\bpublisher{Curran Associates, Inc.}
\end{bincollection}
\endbibitem

\bibitem[\protect\citeauthoryear{Lo}{1984}]{lo;84}
\begin{barticle}[author]
\bauthor{\bsnm{Lo},~\bfnm{A~Y}\binits{A.~Y.}}
(\byear{1984}).
\btitle{On a class of {B}ayesian nonparametric estimates {I}: Density
  estimates}.
\bjournal{The Annals of Statistics}
\bvolume{12}
\bpages{351--357}.
\end{barticle}
\endbibitem

\bibitem[\protect\citeauthoryear{MacEachern}{1999}]{maceachern;99}
\begin{binproceedings}[author]
\bauthor{\bsnm{MacEachern},~\bfnm{S~N}\binits{S.~N.}}
(\byear{1999}).
\btitle{{Dependent nonparametric processes}}.
In \bbooktitle{ASA Proceedings of the Section on Bayesian Statistical Science,
  Alexandria, VA}.
\bpublisher{American Statistical Association}.
\end{binproceedings}
\endbibitem

\bibitem[\protect\citeauthoryear{MacEachern}{2000}]{maceachern;2000}
\begin{btechreport}[author]
\bauthor{\bsnm{MacEachern},~\bfnm{S~N}\binits{S.~N.}}
(\byear{2000}).
\btitle{{Dependent Dirichlet processes}}
\btype{Technical Report},
\bpublisher{Department of Statistics, The Ohio State University}.
\end{btechreport}
\endbibitem

\bibitem[\protect\citeauthoryear{Mena, Ruggiero and
  Walker}{2011}]{mena;ruggiero;walker;2011}
\begin{barticle}[author]
\bauthor{\bsnm{Mena},~\bfnm{R~H}\binits{R.~H.}},
  \bauthor{\bsnm{Ruggiero},~\bfnm{M}\binits{M.}} \AND
  \bauthor{\bsnm{Walker},~\bfnm{S~G}\binits{S.~G.}}
(\byear{2011}).
\btitle{{Geometric stick-breaking processes for continuous-time Bayesian
  nonparametric modeling}}.
\bjournal{Journal of Statistical Planning and Inference}
\bvolume{141}
\bpages{3217--3230}.
\end{barticle}
\endbibitem

\bibitem[\protect\citeauthoryear{Mena and Ruggiero}{2016}]{mena2016}
\begin{barticle}[author]
\bauthor{\bsnm{Mena},~\bfnm{Rams\'es~H.}\binits{R.~H.}} \AND
  \bauthor{\bsnm{Ruggiero},~\bfnm{Matteo}\binits{M.}}
(\byear{2016}).
\btitle{Dynamic density estimation with diffusive Dirichlet mixtures}.
\bjournal{Bernoulli}
\bvolume{22}
\bpages{901--926}.
\bdoi{10.3150/14-BEJ681}
\end{barticle}
\endbibitem

\bibitem[\protect\citeauthoryear{Mira and Petrone}{1996}]{mira;petrone;96}
\begin{binproceedings}[author]
\bauthor{\bsnm{Mira},~\bfnm{A}\binits{A.}} \AND
  \bauthor{\bsnm{Petrone},~\bfnm{S}\binits{S.}}
(\byear{1996}).
\btitle{{Bayesian hierarchical nonparametric inference for change-point
  problems}}.
In \bbooktitle{Bayesian Statistics 5}
(\beditor{\bfnm{J~M}\binits{J.~M.}~\bsnm{Bernardo}},
  \beditor{\bfnm{J~O}\binits{J.~O.}~\bsnm{Berger}},
  \beditor{\bfnm{A~P}\binits{A.~P.}~\bsnm{Dawid}} \AND
  \beditor{\bfnm{A~F~M}\binits{A.~F.~M.}~\bsnm{Smith}}, eds.).
\bpublisher{Oxford University Press}.
\end{binproceedings}
\endbibitem

\bibitem[\protect\citeauthoryear{Muliere and
  Petrone}{1993}]{muliere;petrone;93}
\begin{barticle}[author]
\bauthor{\bsnm{Muliere},~\bfnm{P}\binits{P.}} \AND
  \bauthor{\bsnm{Petrone},~\bfnm{S}\binits{S.}}
(\byear{1993}).
\btitle{{A Bayesian predictive approach to sequential search for an optimal
  dose: parametric and nonparametric models.}}
\bjournal{Journal of the Italian Statistical Society}
\bvolume{2}
\bpages{349--364}.
\end{barticle}
\endbibitem

\bibitem[\protect\citeauthoryear{M\"{u}ller, Erkanli and
  West}{1996}]{mueller;erkanli;west;1996}
\begin{barticle}[author]
\bauthor{\bsnm{M\"{u}ller},~\bfnm{P}\binits{P.}},
  \bauthor{\bsnm{Erkanli},~\bfnm{A}\binits{A.}} \AND
  \bauthor{\bsnm{West},~\bfnm{M}\binits{M.}}
(\byear{1996}).
\btitle{Bayesian curve fitting using multivariate normal mixtures}.
\bjournal{Biometrika}
\bvolume{83}
\bpages{67--79}.
\end{barticle}
\endbibitem

\bibitem[\protect\citeauthoryear{M\"{u}ller, Quintana and
  Rosner}{2004}]{mueller;quintana;rosner;2004}
\begin{barticle}[author]
\bauthor{\bsnm{M\"{u}ller},~\bfnm{P}\binits{P.}},
  \bauthor{\bsnm{Quintana},~\bfnm{F~A}\binits{F.~A.}} \AND
  \bauthor{\bsnm{Rosner},~\bfnm{G}\binits{G.}}
(\byear{2004}).
\btitle{{A method for combining inference across related nonparametric Bayesian
  models}}.
\bjournal{Journal of the Royal Statistical Society, Series B}
\bvolume{66}
\bpages{735--749}.
\end{barticle}
\endbibitem

\bibitem[\protect\citeauthoryear{M\"{u}ller and
  Quintana}{2010}]{mueller;quintana;2010}
\begin{barticle}[author]
\bauthor{\bsnm{M\"{u}ller},~\bfnm{P}\binits{P.}} \AND
  \bauthor{\bsnm{Quintana},~\bfnm{F~A}\binits{F.~A.}}
(\byear{2010}).
\btitle{Random partition models with regression on covariates}.
\bjournal{Journal of Statistical Planning and Inference}
\bvolume{140}
\bpages{2801--2808}.
\end{barticle}
\endbibitem

\bibitem[\protect\citeauthoryear{M\"{u}ller, Quintana and
  Rosner}{2011}]{mueller;quintana;rosner;2011}
\begin{barticle}[author]
\bauthor{\bsnm{M\"{u}ller},~\bfnm{P}\binits{P.}},
  \bauthor{\bsnm{Quintana},~\bfnm{F~A}\binits{F.~A.}} \AND
  \bauthor{\bsnm{Rosner},~\bfnm{G~L}\binits{G.~L.}}
(\byear{2011}).
\btitle{A product partition model with regression on covariates}.
\bjournal{Journal of Computational and Graphical Statistics}
\bvolume{20}
\bpages{260--278}.
\end{barticle}
\endbibitem

\bibitem[\protect\citeauthoryear{M\"{u}ller
  et~al.}{2015}]{mueller;quintana;jara;hanson;2015}
\begin{bbook}[author]
\bauthor{\bsnm{M\"{u}ller},~\bfnm{P}\binits{P.}},
  \bauthor{\bsnm{Quintana},~\bfnm{F~A}\binits{F.~A.}},
  \bauthor{\bsnm{Jara},~\bfnm{A}\binits{A.}} \AND
  \bauthor{\bsnm{E},~\bfnm{Hanson~T}\binits{H.~T.}}
(\byear{2015}).
\btitle{Bayesian Nonparametric Data Analysis}.
\bpublisher{Springer}, \baddress{New York, USA}.
\end{bbook}
\endbibitem

\bibitem[\protect\citeauthoryear{Pr\"unster and Ruggiero}{2013}]{pruenster2013}
\begin{barticle}[author]
\bauthor{\bsnm{Pr\"unster},~\bfnm{Igor}\binits{I.}} \AND
  \bauthor{\bsnm{Ruggiero},~\bfnm{Matteo}\binits{M.}}
(\byear{2013}).
\btitle{A Bayesian nonparametric approach to modeling market share dynamics}.
\bjournal{Bernoulli}
\bvolume{19}
\bpages{64--92}.
\bdoi{10.3150/11-BEJ392}
\end{barticle}
\endbibitem

\bibitem[\protect\citeauthoryear{Quintana and
  Iglesias}{2003}]{quintana;iglesias;2003}
\begin{barticle}[author]
\bauthor{\bsnm{Quintana},~\bfnm{F~A}\binits{F.~A.}} \AND
  \bauthor{\bsnm{Iglesias},~\bfnm{P~L}\binits{P.~L.}}
(\byear{2003}).
\btitle{{Bayesian clustering and product partition models}}.
\bjournal{Journal of The Royal Statistical Society Series B}
\bvolume{65}
\bpages{557--574}.
\end{barticle}
\endbibitem

\bibitem[\protect\citeauthoryear{Regazzini, Lijoi and
  Pr\"{u}nster}{2003}]{regazzini;etal;2003}
\begin{barticle}[author]
\bauthor{\bsnm{Regazzini},~\bfnm{E}\binits{E.}},
  \bauthor{\bsnm{Lijoi},~\bfnm{A}\binits{A.}} \AND
  \bauthor{\bsnm{Pr\"{u}nster},~\bfnm{I}\binits{I.}}
(\byear{2003}).
\btitle{Distributional results for means of normalized random measures with
  independent increments}.
\bjournal{The Annals of Statistics}
\bvolume{31}
\bpages{560--585}.
\end{barticle}
\endbibitem

\bibitem[\protect\citeauthoryear{Reich and Fuentes}{2007}]{reich&fuentes:07}
\begin{barticle}[author]
\bauthor{\bsnm{Reich},~\bfnm{Brian~J.}\binits{B.~J.}} \AND
  \bauthor{\bsnm{Fuentes},~\bfnm{Montserrat}\binits{M.}}
(\byear{2007}).
\btitle{A multivariate semiparametric {B}ayesian spatial modeling framework for
  hurricane surface wind fields}.
\bjournal{The Annals of Applied Statistics}
\bvolume{1}
\bpages{249--264}.
\end{barticle}
\endbibitem

\bibitem[\protect\citeauthoryear{Reinhart}{2019}]{pdfetch}
\begin{bmanual}[author]
\bauthor{\bsnm{Reinhart},~\bfnm{Abiel}\binits{A.}}
(\byear{2019}).
\btitle{pdfetch: Fetch Economic and Financial Time Series Data from Public
  Sources}
\bnote{R package version 0.2.4}.
\end{bmanual}
\endbibitem

\bibitem[\protect\citeauthoryear{Ren et~al.}{2011}]{ren;du;carin;dunson;2011}
\begin{barticle}[author]
\bauthor{\bsnm{Ren},~\bfnm{L}\binits{L.}},
  \bauthor{\bsnm{Du},~\bfnm{L}\binits{L.}},
  \bauthor{\bsnm{Carin},~\bfnm{L}\binits{L.}} \AND
  \bauthor{\bsnm{Dunson},~\bfnm{D~B}\binits{D.~B.}}
(\byear{2011}).
\btitle{Logistic stick-breaking process}.
\bjournal{Journal of Machine Learning Research}
\bvolume{12}
\bpages{203--239}.
\end{barticle}
\endbibitem

\bibitem[\protect\citeauthoryear{Rigon and Durante}{2020}]{RIGON2020}
\begin{barticle}[author]
\bauthor{\bsnm{Rigon},~\bfnm{Tommaso}\binits{T.}} \AND
  \bauthor{\bsnm{Durante},~\bfnm{Daniele}\binits{D.}}
(\byear{2020}).
\btitle{Tractable Bayesian density regression via logit stick-breaking priors}.
\bjournal{Journal of Statistical Planning and Inference}
\bpages{(To appear)}.
\end{barticle}
\endbibitem

\bibitem[\protect\citeauthoryear{Rodr\'{\i}guez, Dunson and
  Gelfand}{2008}]{rodriguez;dunson;gelfand;2008}
\begin{barticle}[author]
\bauthor{\bsnm{Rodr\'{\i}guez},~\bfnm{A}\binits{A.}},
  \bauthor{\bsnm{Dunson},~\bfnm{D~B}\binits{D.~B.}} \AND
  \bauthor{\bsnm{Gelfand},~\bfnm{A}\binits{A.}}
(\byear{2008}).
\btitle{{The nested Dirichlet process}}.
\bjournal{Journal of the American Statistical Association}
\bvolume{103}
\bpages{1131--1154}.
\end{barticle}
\endbibitem

\bibitem[\protect\citeauthoryear{Rodr\'{\i}guez and
  Dunson}{2011}]{rodriguez;dunson;2011}
\begin{barticle}[author]
\bauthor{\bsnm{Rodr\'{\i}guez},~\bfnm{A}\binits{A.}} \AND
  \bauthor{\bsnm{Dunson},~\bfnm{D~B}\binits{D.~B.}}
(\byear{2011}).
\btitle{{Nonparametric Bayesian models through probit stick-breaking
  processes}}.
\bjournal{Bayesian Analysis}
\bvolume{6}
\bpages{145--178}.
\end{barticle}
\endbibitem

\bibitem[\protect\citeauthoryear{Rodr\'{\i}guez and ter
  Horst}{2008}]{rodriguez-terhorst:08}
\begin{barticle}[author]
\bauthor{\bsnm{Rodr\'{\i}guez},~\bfnm{Abel}\binits{A.}} \AND
  \bauthor{\bparticle{ter} \bsnm{Horst},~\bfnm{Enrique}\binits{E.}}
(\byear{2008}).
\btitle{Bayesian dynamic density estimation}.
\bjournal{Bayesian Anal.}
\bvolume{3}
\bpages{339--365}.
\end{barticle}
\endbibitem

\bibitem[\protect\citeauthoryear{Sethuraman}{1994}]{sethuraman;94}
\begin{barticle}[author]
\bauthor{\bsnm{Sethuraman},~\bfnm{J}\binits{J.}}
(\byear{1994}).
\btitle{A constructive definition of {D}irichlet prior}.
\bjournal{Statistica Sinica}
\bvolume{2}
\bpages{639--650}.
\end{barticle}
\endbibitem

\bibitem[\protect\citeauthoryear{Sklar}{1959}]{sklar;59}
\begin{barticle}[author]
\bauthor{\bsnm{Sklar},~\bfnm{A}\binits{A.}}
(\byear{1959}).
\btitle{Fonctions de r\'epartition \`a n dimensions et leurs marges}.
\bjournal{Publications de l'Institut de Statistique de L'Universit\'e d\ e
  Paris}
\bvolume{8}
\bpages{229--231}.
\end{barticle}
\endbibitem

\bibitem[\protect\citeauthoryear{Teh et~al.}{2006}]{teh;jordan;beal;blei;2006}
\begin{barticle}[author]
\bauthor{\bsnm{Teh},~\bfnm{Y~W}\binits{Y.~W.}},
  \bauthor{\bsnm{Jordan},~\bfnm{M~I}\binits{M.~I.}},
  \bauthor{\bsnm{Beal},~\bfnm{M~J}\binits{M.~J.}} \AND
  \bauthor{\bsnm{Blei},~\bfnm{D~M}\binits{D.~M.}}
(\byear{2006}).
\btitle{{Hierarchical Dirichlet processes}}.
\bjournal{Journal of the American Statistical Association}
\bvolume{101}
\bpages{1566--1581}.
\end{barticle}
\endbibitem

\bibitem[\protect\citeauthoryear{Tokdar, Zhu and
  Ghosh}{2010}]{tokdar;zhu;ghosh;2010}
\begin{barticle}[author]
\bauthor{\bsnm{Tokdar},~\bfnm{S~T}\binits{S.~T.}},
  \bauthor{\bsnm{Zhu},~\bfnm{Y~M}\binits{Y.~M.}} \AND
  \bauthor{\bsnm{Ghosh},~\bfnm{J~K}\binits{J.~K.}}
(\byear{2010}).
\btitle{{Bayesian density regression with logistic Gaussian process and
  subspace projection}}.
\bjournal{Bayesian Analysis}
\bvolume{5}
\bpages{1--26}.
\end{barticle}
\endbibitem

\bibitem[\protect\citeauthoryear{Trippa, M\"uller and Johnson}{2011}]{Trippa11}
\begin{barticle}[author]
\bauthor{\bsnm{Trippa},~\bfnm{Lorenzo}\binits{L.}},
  \bauthor{\bsnm{M\"uller},~\bfnm{Peter}\binits{P.}} \AND
  \bauthor{\bsnm{Johnson},~\bfnm{Wesley}\binits{W.}}
(\byear{2011}).
\btitle{The multivariate beta process and an extension of the {P}olya tree
  model}.
\bjournal{Biometrika}
\bvolume{98}
\bpages{17-34}.
\end{barticle}
\endbibitem

\bibitem[\protect\citeauthoryear{Wade, Mongelluzzo and
  Petrone}{2011}]{wade2011}
\begin{barticle}[author]
\bauthor{\bsnm{Wade},~\bfnm{Sara}\binits{S.}},
  \bauthor{\bsnm{Mongelluzzo},~\bfnm{Silvia}\binits{S.}} \AND
  \bauthor{\bsnm{Petrone},~\bfnm{Sonia}\binits{S.}}
(\byear{2011}).
\btitle{An enriched conjugate prior for Bayesian nonparametric inference}.
\bjournal{Bayesian Anal.}
\bvolume{6}
\bpages{359--385}.
\bdoi{10.1214/11-BA614}
\end{barticle}
\endbibitem

\bibitem[\protect\citeauthoryear{Wang and Rosner}{2019}]{wang&rosner:19}
\begin{barticle}[author]
\bauthor{\bsnm{Wang},~\bfnm{Chenguang}\binits{C.}} \AND
  \bauthor{\bsnm{Rosner},~\bfnm{Gary~L.}\binits{G.~L.}}
(\byear{2019}).
\btitle{A Bayesian nonparametric causal inference model for synthesizing
  randomized clinical trial and real-world evidence}.
\bjournal{Statistics in Medicine}
\bvolume{38}
\bpages{2573-2588}.
\bdoi{10.1002/sim.8134}
\end{barticle}
\endbibitem

\bibitem[\protect\citeauthoryear{Xu, MacEachern and
  Xu}{2015}]{xu;maceachern;xu;15}
\begin{barticle}[author]
\bauthor{\bsnm{Xu},~\bfnm{Z.}\binits{Z.}},
  \bauthor{\bsnm{MacEachern},~\bfnm{S.~N.}\binits{S.~N.}} \AND
  \bauthor{\bsnm{Xu},~\bfnm{X.}\binits{X.}}
(\byear{2015}).
\btitle{Modeling non-{G}aussian time series with nonparametric {B}ayesian
  model}.
\bjournal{IEEE Transactions on Pattern Analysis and Machine Intelligence}
\bvolume{37}
\bpages{372--382}.
\end{barticle}
\endbibitem

\end{thebibliography}

\end{document}